\documentclass[manuscript,screen]{acmart}
\AtBeginDocument{%
  }

\setcopyright{acmlicensed}
\copyrightyear{2018}
\acmYear{2018}
\acmDOI{XXXXXXX.XXXXXXX}
\begin{document}

\title{Patterns, Models, and Challenges in Online Social Media: A Survey}

\author{Niccolò Di Marco}
\email{niccolo.dimarco@unitus.it}
\orcid{0000-0003-4335-7328}
\affiliation{%
  \institution{University of Tuscia}
  \city{Viterbo}
  \country{Italy}
}

\author{Anita Bonetti}
\email{anita.bonetti@uniroma1.it}
\orcid{0009-0002-9507-7379}
\affiliation{%
  \institution{Sapienza University of Rome}
  \city{Rome}
  \country{Italy}
}

\author{Edoardo Di Martino}
\email{edoardo.dimartino@uniroma1.it}
\orcid{0009-0000-8301-136X}
\affiliation{%
  \institution{Sapienza University of Rome}
  \city{Rome}
  \country{Italy}
}

\author{Edoardo Loru}
\email{edoardo.loru@uniroma1.it}
\orcid{0009-0007-4629-930X}
\affiliation{%
  \institution{Sapienza University of Rome}
  \city{Rome}
  \country{Italy}
}

\author{Jacopo Nudo}
\email{jacopo.nudo@uniroma1.it}
\orcid{0009-0003-7169-0774}
\affiliation{%
  \institution{Sapienza University of Rome}
  \city{Rome}
  \country{Italy}
}

\author{Mario Edoardo Pandolfo}
\email{marioedoardo.pandolfo@uniroma1.it}
\orcid{0009-0006-6509-4425}
\affiliation{%
  \institution{Sapienza University of Rome}
  \city{Rome}
  \country{Italy}
}

\author{Giulio Pecile}
\email{giulio.pegile@uniroma1.it}
\orcid{0009-0004-0288-2542}
\affiliation{%
  \institution{Sapienza University of Rome}
  \city{Rome}
  \country{Italy}
}

\author{Emanuele Sangiorgio}
\email{emanuele.sangiorgio@uniroma1.it}
\orcid{0009-0003-1024-3735}
\affiliation{%
  \institution{Sapienza University of Rome}
  \city{Rome}
  \country{Italy}
}

\author{Irene Scalco}
\email{irene.scalco@uniroma1.it}
\orcid{0009-0007-9151-0303}
\affiliation{%
  \institution{Sapienza University of Rome}
  \city{Rome}
  \country{Italy}
}

%%%%
\author{Simon Zollo}
\email{simon.zollo@uniroma1.it}
\orcid{0009-0004-0682-8578}
\affiliation{%
  \institution{Sapienza University of Rome}
  \city{Rome}
  \country{Italy}
}

\author{Matteo Cinelli}
\orcid{0000-0003-3899-4592}
\email{matteo.cinelli@uniroma1.it}
\affiliation{%
  \institution{Sapienza University of Rome}
  \city{Rome}
  \country{Italy}
}

\author{Fabiana Zollo}
\orcid{0000-0002-0833-5388}
\email{fabiana.zollo@unive.it}
\affiliation{%
  \institution{Ca' Foscari University of Venice}
  \city{Venice}
  \country{Italy}
}

\author{Walter Quattrociocchi}
\orcid{0000-0002-4374-9324}
\email{walter.quattrociocchi@uniroma1.it}
\affiliation{%
  \institution{Sapienza University of Rome}
  \city{Rome}
  \country{Italy}
}

\renewcommand{\shortauthors}{Di Marco et al.}

\begin{abstract}
The rise of digital platforms has enabled the large-scale observation of individual and collective behavior through high-resolution interaction data. This development has opened new analytical pathways for investigating how information circulates, how opinions evolve, and how coordination emerges in online environments. Yet despite a growing body of research, the field remains fragmented—marked by methodological heterogeneity, limited model validation, and weak integration across domains.
This survey offers a systematic synthesis of empirical findings and formal models. We examine platform-level regularities, assess the methodological architectures that generate them, and evaluate the extent to which current modeling frameworks account for observed dynamics.
The goal is to consolidate a shared empirical baseline and clarify the structural constraints that shape inference in this domain, laying the groundwork for more robust, comparable, and actionable analyses of online social systems.
\end{abstract}

\begin{CCSXML}
<ccs2012>
<concept>
<concept_id>10010405.10010455.10010461</concept_id>
<concept_desc>Applied computing~Sociology</concept_desc>
<concept_significance>300</concept_significance>
</concept>
<concept>
<concept_id>10002951.10003260</concept_id>
<concept_desc>Information systems~World Wide Web</concept_desc>
<concept_significance>100</concept_significance>
</concept>
<concept>
<concept_id>10002978.10003029</concept_id>
<concept_desc>Security and privacy~Human and societal aspects of security and privacy</concept_desc>
<concept_significance>100</concept_significance>
</concept>
<concept>
<concept_id>10010405</concept_id>
<concept_desc>Applied computing</concept_desc>
<concept_significance>300</concept_significance>
</concept>
</ccs2012>
\end{CCSXML}

\ccsdesc[300]{Applied computing~Sociology}
\ccsdesc[100]{Information systems~World Wide Web}
\ccsdesc[100]{Security and privacy~Human and societal aspects of security and privacy}
\ccsdesc[300]{Applied computing}

\keywords{Social media, online behaviour, computational social science, data science}

\maketitle

\section{Introduction}
The widespread adoption of digital platforms has profoundly reshaped how individuals access information \cite{del2016spreading,vosoughi2018spread}, interact with content \cite{schmidt2017anatomy,kumpel2015news,tsagkias2011linking}, and engage in public discourse \cite{centola2010spread,johnson2019hidden}. Social media platforms, in particular, mediate these interactions through algorithmic personalization \cite{guy2010social}, content ranking \cite{momeni2015survey}, and engagement-based feedback loops \cite{gillespie2018custodians}. As a result, they have become not only channels of communication, but also active components in the structure of opinion formation, public debate, and collective behavior \cite{Bail2018,tucker2018social}.

Empirical research over the past decade has identified several recurring behavioral patterns associated with these systems \cite{injadat2016data}. Users tend to engage more with content that confirms their pre-existing views~\cite{del2016spreading,flaxman2016filter,guess2018selective,budak2024misunderstanding}, contributing to ideological polarization and reducing exposure to alternative perspectives~\cite{bessi2015science,cinelli2021echo,garimella2018political,barbera2015tweeting,del2017mapping}.
False or misleading content can achieve wide circulation, often outperforming accurate information in terms of reach and engagement~\cite{del2016spreading,vosoughi2018spread}. These dynamics have been linked to downstream effects in diverse domains, including public health~\cite{cinelli2020covid,do2022infodemics,alvarez2021determinants,bashir2017effects,pierri2022online,burki2019vaccine}, mental health~\cite{Orben2022,Naslund2016,bashir2017effects,weigle2024social,ferguson2024there}, and political communication~\cite{pecile2025mapping,zhuravskaya2020political,tucker2018social,metaxas2012social,aral2019protecting}.

A key factor enabling observations has been the increasing availability of large-scale digital behavioral data~\cite{ruths2014social,lazer2020computational}. In fact, logs of user interactions—such as likes, shares, and comments- have provided direct, time-stamped evidence of platform-mediated behavior at scale~\cite{gonzalez2011dynamics,Avalle2024,wagner2021measuring,ruths2014social}.

This has led to a shift in the methodological foundations of social research: from small-scale, self-reported, or experimental studies toward empirical analyses grounded in observed behavior. This transition has catalyzed the growth of computational social science~\cite{lazer2009computational,Conte2012,watts2007twenty,gonzalez2016networked}, an interdisciplinary field that applies methods from network science, machine learning, and statistics to study social phenomena through digital traces.

However, the emergence of behavioral data as the main source of evidence has also introduced new challenges~\cite{lazer2020computational}. While it enables the detection of reproducible patterns and statistical regularities, it provides limited access to underlying cognitive processes or motivations~\cite{mosleh2020self}. Observational data are also subject to structural biases and rarely allow for causal inference in the absence of experimental variation~\cite{tufekci2014big}. 

Moreover, as research has expanded, the field has become increasingly fragmented~\cite{wang2023less}. Separate literatures have developed around specific domains—such as polarization, misinformation, attention, algorithmic exposure, and coordination—each with distinct assumptions, datasets, and methodologies~\cite{paakkonen2024emergence,pablo2014social,lelkes2017hostile,vicario2019polarization}. This fragmentation has limited the comparability of results across studies and platforms, and has made it difficult to build a cumulative understanding of online social dynamics~\cite{freelon2018computational}. 

These limitations suggest the need for a comparative and conceptually integrated approach. Much of the existing literature remains tied to individual platforms, making it difficult to determine whether observed effects are driven by human behavior, content properties, or platform-specific design~\cite{guess2020misinformation,wagner2021measuring}. 
Only recent works have begun to adopt multiplatform and longitudinal strategies to address this limitation~\cite{Avalle2024, DiMarco2024}. These approaches provide important insights into which patterns are system-specific and which are more general, offering a stronger basis for explanation, prediction, and intervention.

The present survey responds to the above limitations by offering a structured synthesis of current research on online behavior from a data science perspective. Our aim is to provide a coherent analysis of the development of the field, establishing the basis for future research. 

The paper is organized into these sections:
Section \ref{sec:shift} examines the origins of computational social science, highlighting the pivotal shift from self-reported experiments to big data approaches.
Section \ref{sec:methods} outlines the data sources and empirical methods commonly used to study online behavior.
Section \ref{sec:key_phenomena} synthesizes key phenomena uncovered through large-scale analyses, including selective exposure, agenda-setting dynamics, echo chambers, misinformation, and coordinated behavior.
Section \ref{sec:opinion_dynamics} presents a brief history of opinion dynamics models, focusing on the most influential approaches.

Building on these foundations, Section \ref{sec:comparative} highlights the need for a comparative and longitudinal perspective, discussing open challenges and future directions toward a cumulative, comparative science of digital behavior.

Finally, we discuss the current limitations, open research directions, and design implications in Sections \ref{sec:limitations} and \ref{sec:design}.
We conclude the paper by summarizing the key findings in Section \ref{sec:conclusion}.

\section{The Shift Toward Behavioral Data}\label{sec:shift}

Social behavior studies have traditionally relied on controlled experiments, small-sample surveys, and self-reported measures \cite{carpenter2011evaluating,reid2004some,tsui1997empirical}. While these methods have produced important theoretical insights, they face well-documented limitations \cite{bell2012obstacles}. Experimental findings often struggle to generalize beyond specific contexts~\cite{Button2013,Makel2012}, and self-reports are affected by recall errors, framing effects, and social desirability biases~\cite{Furnham1986,Rosenman2011,Koller2023}. As a result, key aspects of online behavior—such as attention allocation, content exposure, and interaction dynamics—have long remained difficult to observe directly and with sufficient scale.

The emergence of digital platforms has introduced a structural change in how human behavior can be measured \cite{wu2021platform}. User actions—such as likes, comments, shares, and clicks—generate large volumes of time-stamped, structured data \cite{kim2017like,schreiner2021impact}. These digital traces capture in-situ behavior unfolding in real-world conditions and at population scale~\cite{Bell2009,Parry2021,sultan2023leaving}.
This shift from declarative to behavioral evidence has redefined the empirical foundations of social research.

From a methodological perspective, the impact is twofold. First, digital data allow for large-scale, longitudinal observation across millions of users, enabling the detection of macro-level regularities that were previously inaccessible \cite{kumar2011understanding,adedoyin2014survey}. Second, they support diagnostic analysis of platform-mediated processes, such as information diffusion \cite{li2017survey,guille2013information}, community formation \cite{backstrom2006group,del2016echo}, and feedback amplification \cite{lim2022opinion,huszar2022algorithmic}. These phenomena are dynamic and shaped by technical, cognitive, and social constraints, making them difficult to study through traditional methods alone.

This transition has contributed to the emergence of computational social science \cite{lazer2009computational,Conte2012}, an interdisciplinary field integrating network analysis, machine learning, and statistical inference to study social dynamics through digital evidence. The core premise is that online platforms generate behavioral signals that are both ecologically valid and methodologically rich, offering a new basis for empirical inquiry into collective behavior.
However, the quest remains highly fragmented with both challenges and opportunities \cite{lazer2020computational}.

\begin{figure}[!ht]
\centering
\includegraphics[width=0.5\linewidth]{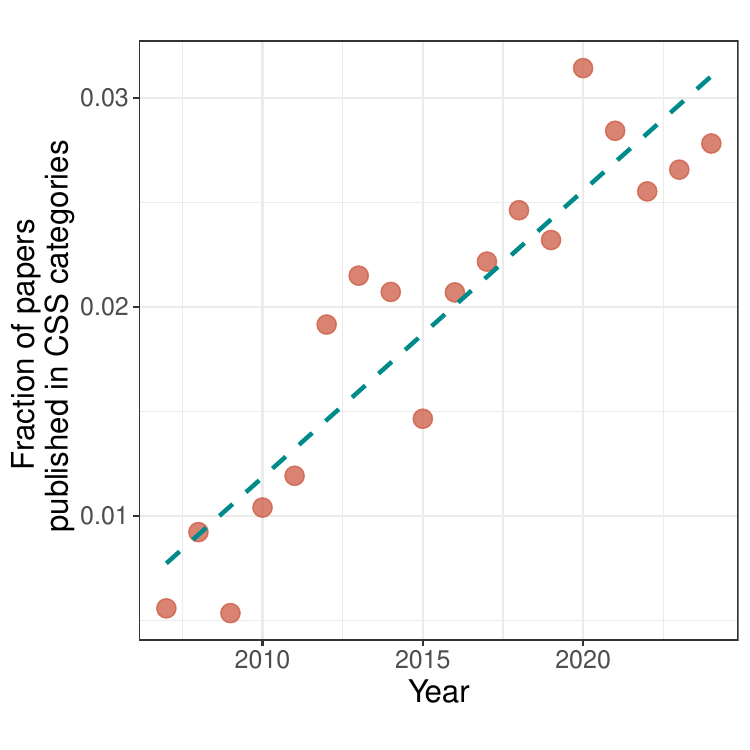}
\caption{Fraction of papers published on arXiv in cs.CY, cs.SI, and physics.soc-ph — three key categories in computational social science (CSS). The notable increase highlights the growing influence of digital behavioral data and computational methods in the empirical study of online social dynamics.}
\label{fig:arxiv}
\end{figure}

To illustrate this shift, Figure~\ref{fig:arxiv} presents the fraction of papers published on arXiv between 2007 and 2024 in three relevant categories: Computers and Society (cs.CY), Social and Information Networks (cs.SI), and Physics and Society (physics.soc-ph).
A notable increase in submissions is evident, coinciding with the broader availability of web-scale behavioral data. This growth reflects expanded research capacity and a fundamental change in epistemological orientation: from hypothesis-driven studies based on small samples to large-scale analyses grounded in direct observation.

However, this new empirical regime also introduces important challenges. Access to behavioral data is often restricted due to proprietary controls or lack of standardized APIs \cite{acker2020social}. 

Moreover, the abundance of data does not guarantee explanatory clarity: observational signals require careful methodological treatment to rule out confounding factors and support causal inference. In addition, digital behavior is not neutral: the design of interfaces, algorithmic ranking systems, and platform affordances shapes it \cite{abrams2025neutrality,hallinan2022beyond}. This complicates the task of disentangling endogenous user preferences from exogenous system effects \cite{Avalle2024}.

Despite its limitations, the rise of behavioral data marks a significant turning point in research \cite{ruths2014social, zafarani2014behavior}. It opens up new research questions, validation strategies, and applications in system design, content governance, and policy evaluation. Rather than replacing traditional approaches, it reconfigures them, shifting the emphasis from interpretation to structural regularities, and from isolated experiments to population-level diagnostics. The following sections explore how this shift unfolds across various methodological strategies and research domains.

\section{Methodological Foundations}
\label{sec:methods}

The study of online social behavior relies on a diverse set of empirical strategies, each offering access to different aspects of digital dynamics. These approaches—ranging from large-scale observation to controlled experimentation and computational modeling—have individually produced valuable insights. However, their integration remains limited \cite{cohen2013classifying,jurgens2015geolocation}. As a result, the field is marked by fragmentation: different methods often lead to non-comparable results, and cumulative knowledge remains difficult to achieve \cite{paakkonen2024emergence,wang2023less}.

To move beyond isolated findings, a consolidated methodological framework is needed—one that makes explicit the assumptions, capabilities, and structural limitations of each approach. In this section, we review the principal empirical strategies used to investigate phenomena such as polarization, algorithmic exposure, misinformation, and coordinated behavior. Rather than aiming for taxonomic completeness, our goal is conceptual clarity: to highlight what each method reveals, what it obscures, and how triangulation across methods can mitigate blind spots.

\subsection{Empirical Strategies: Strengths and Structural Constraints}

Research on platform-mediated behavior typically draws on six broad methodological families, each with distinct affordances and limitations. To shed light on the internal dynamics of each approach, this section is entirely dedicated to their discussion. Table \ref{tab:methods} summarizes the approaches.

\textbf{Observational network analyses} use digital traces—such as follower graphs \cite{bovet2019influence,flamino2023political}, sharing cascades \cite{del2016spreading,weng2013virality,wagner2021measuring}, or hyperlink structures—to map large-scale patterns of interaction and content flow \cite{schmidt2017anatomy,johnson2020online,park2003hyperlink}. These methods are fundamental for detecting emergent regularities (see Section~\ref{sec:attention}), but are limited to correlational inference. They often suffer from visibility bias and provide little information about actual exposure or user-level interpretation.

\textbf{Field experiments}, including randomized interventions on algorithms \cite{bandy2023facebook,oeldorf2020ineffectiveness,martel2024fact,lazer2015rise}, friction elements, or content labels, allow for causal identification of treatment effects in real-world settings \cite{bandy2023facebook,oeldorf2020ineffectiveness,martel2024fact,lazer2015rise}. These studies offer high external validity and practical relevance (e.g., for agenda-setting dynamics, Section~\ref{sec:agenda}), but typically depend on privileged access to platform infrastructure. This restricts both transparency and reproducibility.

\textbf{Survey and laboratory experiments} enable precise measurement of beliefs, attitudes, and perception under controlled conditions \cite{lewandowsky2021countering,mosleh2022field}. These methods are strong in internal validity and useful for theory testing. However, these studies often rely on small or skewed samples, and their experiments are typically conducted in simulated—and therefore simplified—scenarios. As a result, their relevance to real platform behavior remains limited, especially in contexts involving coordinated activity (Section~\ref{sec:coordination}).

\textbf{Algorithmic audits} simulate user behavior via bots or shadow accounts to probe the logic of content recommendation systems \cite{bartley2021auditing,shen2021everyday,bandy2021problematic,scott2011auditing}. These methods are valuable for identifying structural biases in exposure (see Section~\ref{sec:algorithmic}), but face limits in extrapolating from synthetic agents to real users. In some cases, platform terms of service constrain their implementation or reproducibility.

\textbf{Browser-based and panel instrumentation}, often based on opt-in tracking or user panels, provides fine-grained data on actual exposure and media consumption \cite{goel2012does,kumar2010characterization,bucklin2003model,metzger2018benefits}. These methods are particularly relevant for studying the informational environment surrounding phenomena such as polarization and misinformation (Section~\ref{sec:misinformation}). However, they are vulnerable to selection bias, underrepresentation of mobile usage, and privacy-related dropout.

\textbf{Agent-based and simulation models} formalize assumptions about cognition, influence, and feedback loops to explore systemic outcomes and counterfactuals \cite{rand2015agent,toriumi2012people,plikynas2015agent,kaligotla2015agent}. These models provide mechanistic insight and aid in conceptualizing dynamics under uncertainty. Yet, their explanatory value depends heavily on empirical calibration—an aspect often neglected when real-world data are unavailable or siloed.

\begin{table}[H]
\centering
\caption{Empirical strategies for studying online behavior: scope, structural limits, and typical applications.}
\label{tab:methods}
\renewcommand{\arraystretch}{1.05}
\setlength{\tabcolsep}{2pt}
\scriptsize
\begin{tabular}{|p{3.2cm}|p{3.8cm}|p{3.8cm}|p{2.7cm}|}
\hline
\textbf{Method} & \textbf{Scope} & \textbf{Limitations} & \textbf{Typical Use} \\
\hline
Observational network analysis & Large-scale interaction patterns, content flow, polarization & No causal inference; visibility bias; lacks exposure data & Mapping cascades, detecting echo chambers \\
\hline
Field experiments & Causal effects of ranking changes, labels, friction, exposure modulation & Requires platform collaboration; limited replicability & Testing impact of algorithmic interventions \\
\hline
Surveys and lab experiments & Attitudes, belief shifts, perceived credibility under control & Low external validity; sample skew; disconnect from real behavior & Evaluating fact-checking or inoculation effects \\
\hline
Algorithmic audits & Personalization dynamics, recommendation logic, content visibility & Synthetic behavior; platform ToS limits; hard to generalize & Probing filter bubbles and recommender bias \\
\hline
Browser/panel instrumentation & Real exposure traces, behavior across demographics & Opt-in bias; poor mobile coverage; privacy attrition & Reconstructing media diets and exposure-belief links \\
\hline
Agent-based / simulation models & Mechanisms of influence, polarization, opinion formation & Stylized assumptions; often unvalidated; limited predictive scope & Exploring counterfactuals and systemic dynamics \\
\hline
\end{tabular}
\end{table}

\subsection{Comparative Synthesis and Empirical Trade-offs}
Despite their diversity, the previous strategies face a common set of structural tensions that shape the boundaries of current knowledge.

First, there is a trade-off between \emph{scale and causality}. Large-scale observational studies are well-suited to detect aggregate patterns \cite{Avalle2024, johnson2020online} but typically cannot support causal claims \cite{zollo2024understanding}. Conversely, methods that enable causal inference—such as randomized field experiments—often operate on limited populations, within narrow contexts, or under artificial conditions.

Second, \emph{access asymmetry} remains a major obstacle. The most informative interventions, especially those involving the internal mechanisms of platforms (e.g., feed ranking, interface design), require deep integration with proprietary infrastructures. This restricts both academic independence and reproducibility \cite{wagner2023independence,verma2014editorial, esteve2017business}.

Third, the field is affected by a persistent \emph{validation gap} between theoretical models and empirical observations. Simulation-based approaches offer plausible mechanisms, but are often disconnected from data. Without a systematic grounding in behavioral evidence, models risk internal coherence without real-world relevance \cite{fagiolo2006empirical,manzo2014potentialities,windrum2007empirical}.

These tensions are not incidental. They define the epistemic constraints of the field and delimit what can—and cannot—be inferred about online behavior. Recognizing these structural boundaries is essential for interpreting results, designing robust studies, and building a cumulative science of online social dynamics \cite{ghani2019social, lazer2009computational, Conte2012}.

\section{Key Phenomena in Online Social Dynamics}\label{sec:key_phenomena}

Social platforms have deeply restructured the architecture of public discourse \cite{Avalle2024,kitsak2010identification,corbyn2012facebook}, altering how attention is allocated \cite{menczer2020attention,ciampaglia2015production,goldhaber1997attention} and how content spreads \cite{del2016spreading,pei2014searching,cinelli2020covid,zareie2021minimizing}. 

In the sections that follow, we adopt a dual perspective. We review substantive findings across key domains: attention and exposure (Section~\ref{sec:attention}), agenda-setting (Section~\ref{sec:agenda}), algorithmic influence (Section~\ref{sec:algorithmic}), misinformation (Section~\ref{sec:misinformation}), and coordination (Section~\ref{sec:coordination})—while keeping the methodological foundations in view. 

Together, these processes form the backbone of the socio-technical logic that governs digital information dynamics. 
Our goal is to show not only what the literature reveals, but also how those insights depend on specific empirical architectures, and where future progress is needed.

\subsection{Selective Exposure and Attention Allocation}
\label{sec:attention}

The attention economy plays a central role in structuring online content consumption \cite{an2013fragmented,Davenport2001,Falkinger2008,spohr2017fake}. In a context of chronic information abundance \cite{anderson2012competition, Hargittai2012,Bawden2020}, visibility becomes a limited and contested resource. Rather than emerging from user-driven search and intentional discovery, exposure is increasingly mediated by feed architectures, push-based notification systems, and mobile-first interfaces \cite{cinelli2020selective,Waller2012,Bawden2020}. These affordances prioritize immediacy and reactivity, delivering a continuous stream of stimuli that compete for cognitive bandwidth in real time \cite{Tufekci2013,Bhargava2020}.

Empirical analyses of user interactions on platforms such as Facebook, YouTube, and Twitter consistently show a strong concentration of attention: a small fraction of content receives a disproportionately large share of engagement. The distribution of such engagement typically follows heavy-tailed distributions, remarkably consistent across platforms, content types, and time windows \cite{weng2013virality,bessi2016users,cheng2008statistics}.

This skew is not primarily determined by the size of the audience or the institutional reputation of the content producer. Some studies show that even pages with relatively few followers can outperform established outlets when their content aligns with emotionally salient or identity-relevant narratives \cite{schmidt2017anatomy,Sangiorgio2024,lerman2007social}. 

This concentration becomes even more pronounced during major events, where content that triggers emotional reactions or reinforces group identity tends to attract more engagement than content selected for accuracy or relevance \cite{Etta2023,Robertson2023,torricelli2023does, gallotti2020assessing}.

These patterns may not be driven by user behavior alone. They could be reinforced by platform algorithms that tend to rank content based on engagement metrics—such as likes, shares, comments, or watch time—rather than on indicators of accuracy or credibility \cite{cinelli2021echo}. 

If engagement functions as the main proxy for relevance, a feedback loop may emerge: content that attracts interaction becomes more visible, which in turn increases its chances of spreading \cite{bessi2015viral,bak2022combining,kim2018they}. 

This mechanism may systematically favor posts that are emotional, polarizing, or identity-driven, selecting for content optimized for immediate reactions rather than thoughtful analysis \cite{del2016echo,mousavi2022effective,nikolinakou2018viral}.

Temporal analyses of viral content show that visibility is not just concentrated, but also highly volatile \cite{fernandez2024analysing,guerini2011exploring,han2020importance, di2024users}.
Spikes in attention usually last only a few hours or days, yet they can have lasting effects on how content is ranked and how users perceive what is important \cite{sangiorgio2025evaluating}. 

This volatility creates strong pressure to produce content that captures attention quickly. As a result, platforms may end up favoring posts that are simple, emotionally charged, and repetitive, rather than those that require context, continuity, or fact-checking \cite{di2024users,stieglitz2013emotions,Amini2025}.

The implications may be structural. The way attention is allocated —a key feature of the digital public sphere— appears increasingly influenced by optimization processes that prioritize behavioral signals over informational value \cite{ecker2022psychological,briand2021infodemics}. 
Rather than distorting exposure in isolated instances, these dynamics could be reshaping the default conditions under which content gains visibility, with potential consequences for pluralism, deliberation, and the quality of public information.

\subsection{Agenda Setting and Content Visibility}
\label{sec:agenda}

Agenda setting refers to the process by which certain topics, issues, or narratives gain prominence in public discourse, shaping what people perceive as salient or worthy of attention \cite{mccombs1972agenda}. In classical media theory, this process was conceptualized as largely top-down: editors, broadcasters, and political institutions selected and framed content for mass audiences.

In today’s platform-mediated environments, agenda setting has become decentralized, continuous, and personalized \cite{gilardi2022social,yang2016social}. Content exposure is no longer orchestrated by a limited set of gatekeepers, but emerges from the interaction between algorithmic curation, user engagement patterns, and platform-specific design features \cite{quattrociocchi2014opinion}. This shift alters not only who can influence the public agenda, but also how visibility itself is structured and distributed \cite{kim2025agenda}.

In this environment, feeds, trending sections, and push notifications function as de facto gatekeepers \cite{meraz2009there}. Rather than enforcing editorial hierarchies, these mechanisms react to behavioral signals in real time, adjusting visibility according to measures of popularity, velocity, and interaction density \cite{ding2019social,kong2018exploring}. As a result, the salience of a topic is increasingly determined not by its institutional relevance or civic importance, but by its capacity to trigger engagement at scale \cite{trunfio2021conceptualising,del2016echo}.

Empirical studies suggest that visibility on digital platforms is no longer limited to legacy media or high-follower accounts. Individual users and niche communities —especially when their content aligns with platform engagement dynamics— can reach exposure levels comparable to those of institutional actors \cite{Sangiorgio2024}. Algorithmic ranking systems contribute to this shift by prioritizing interaction metrics, which may weaken the link between visibility and traditional indicators such as source credibility or editorial oversight \cite{zaccaria2019poprank}.

Experimental studies show that agenda dynamics are highly sensitive to even minor changes in platform design. Adjustments to ranking algorithms, friction elements, or interface layout can significantly influence the visibility, reach, and perceived importance of content \cite{moehring2023providing,bakshy2015exposure}. These results indicate that agenda setting is shaped not only by user interest but also by the structural rules through which platforms translate engagement into visibility.

Network-based analyses add an additional layer of complexity \cite{weng2018attention}. Information diffusion often occurs within tightly clustered communities, driven by homophily and shared identity markers \cite{del2016echo,cinelli2021echo,gonccalves2011modeling}. In such contexts, content salience may be amplified not through broad dissemination but through localized resonance. 
In \cite{Loru2024agenda}, the authors show that topics rise to prominence not necessarily because they are imposed from above, but because they activate affective or identity-based responses that trigger rapid, self-reinforcing attention loops within specific user segments \cite{falkenberg2022growing}.

Taken together, these findings raise questions about the continued relevance of hierarchical media models for explaining how content gains visibility online. Agenda setting now appears to operate as a distributed process shaped by feedback loops, algorithmic filtering, and platform-specific design choices. Explaining which topics rise to prominence—and under what conditions—calls for analytical approaches that account for the interplay between user behavior, network structure, and system-level constraints.

\subsection{Algorithmic Amplification and Echo Chambers}
\label{sec:algorithmic}

The mechanisms of agenda setting discussed above are structurally intertwined with algorithmic curation—the process by which digital platforms personalize content exposure based on predicted user engagement \cite{covington2016deep}. These curation systems operate through continuous optimization of ranking algorithms, leveraging behavioral signals such as clicks, shares, watch time, and interaction rate to maximize retention and activity \cite{biega2018equity}. In doing so, they establish feedback loops in which user behavior influences content visibility, which in turn shapes subsequent behavior \cite{bakshy2014designing,quattrociocchi2017inside}. Over time, this dynamic fosters homophily-based reinforcement and the emergence of ideologically homogeneous environments, commonly referred to as echo chambers \cite{cinelli2021echo,mahmoudi2024echo}.

These structures are not explicitly designed but arise endogenously \cite{cinelli2021echo} from the cumulative effects of personalization strategies. As platforms optimize content exposure to maximize short-term engagement, users are increasingly shown material that aligns with their prior preferences, social identity, or emotional disposition \cite{bakshy2015exposure,del2016spreading}. This narrowing of informational inputs reduces exposure diversity and inhibits cross-cutting interaction\cite{bessi2015science,dubois2018echo}, particularly in politically or emotionally charged contexts \cite{falkenberg2024patterns,vicario2019polarization}.

Platforms with high algorithmic curation—such as Facebook, YouTube, and TikTok—consistently exhibit stronger ideological clustering \cite{gaines2009typing}, higher content redundancy, and lower topical diversity than platforms with weaker or absent personalization, such as Reddit or Gab \cite{cinelli2021echo,munger2019supply}. These patterns persist across languages and national contexts \cite{falkenberg2024patterns}, suggesting a structural effect of curation logic rather than content-specific dynamics \cite{hartmann2025systematic}.

Sentiment analyses further indicate that echo chambers are not only informationally homogeneous \cite{amendola2024towards,williams2015network,Zollo2015} but also emotionally unstable \cite{brugnoli2019recursive,del2016echo,sasahara2021social}. Moreover, recent analyses show that content expressing out-group animosity, particularly from political actors, consistently drives higher engagement across platforms, reinforcing the emotional polarization of online discussions \cite{Rathje2021}. As homogeneity increases, conversations within these environments tend to become more negative, reactive, and polarized \cite{Zollo2015}. This affective decay is reinforced by the reward structure of engagement-driven ranking: content that elicits outrage, fear, or in-group affirmation is more likely to be promoted, accelerating the circulation of emotionally charged narratives \cite{del2016echo,choi2020rumor}.

Importantly, viral success is decoupled from traditional metrics of influence such as follower count or institutional authority \cite{Sangiorgio2024,robertson2023negativity,Etta2023}.

This creates a bias toward polarizing information and reduces the visibility of pluralistic or deliberative discourse.

The amplification dynamics induced by algorithmic systems thus constitute a core mechanism in the formation and persistence of echo chambers. Their effects are not limited to exposure filtering but extend to the structuring of emotional tone, the shaping of perceived consensus, and the suppression of informational diversity. Understanding these dynamics is critical for interpreting observed patterns of polarization and for designing interventions that address their systemic roots \cite{cinelli2021echo,del2016spreading}.

\subsection{(Mis)information Dynamics and Polarization}
\label{sec:misinformation}

The dynamics of algorithmic amplification described above contribute directly to the persistence and reach of misinformation in digital ecosystems. In environments optimized for engagement, the diffusion of content is shaped less by its factual accuracy than by its emotional salience, identity alignment, and resonance with existing beliefs \cite{guess2020exposure,bessi2015science}. Large-scale behavioral studies show that users systematically prefer content that confirms their prior views, particularly when embedded in socially coherent communities such as ideological echo chambers \cite{pennycook2019lazy,DelVicario2016,Zollo2017}.

This preference creates favorable conditions for the spread of misinformation \cite{ciampaglia2018research,lazer2018science}. False or misleading claims often outperform accurate information \cite{vosoughi2018spread} in terms of engagement because they tend to be simpler, more emotionally charged, and more aligned with in-group narratives \cite{del2016spreading}. These characteristics make misinformation particularly effective in low-reflection contexts and in feed environments where rapid interaction is incentivized \cite{juul2021comparing}.

Attempts to mitigate misinformation through fact-checking \cite{martel2021you}, warning labels \cite{martel2023misinformation,jia2022understanding}, or inoculation campaigns have produced mixed outcomes \cite{maertens2021long}. A large-scale analysis of over 50,000 debunking posts on Facebook revealed that corrective content mostly circulates within communities already predisposed to accept the correction, with negligible penetration into conspiratorial or misinformed segments\cite{Zollo2017}. In practice, debunking often fails to reach those most exposed to or influenced by misinformation \cite{lewandowsky2012misinformation,iizuka2022impact,chan2023meta}.

Further research suggests that cognitive alignment often takes precedence over informational intent \cite{taylor2023identity}. Users with strong ideological or conspiratorial predispositions may interpret corrections, satire, or disclaimers not as challenges to their beliefs, but as further confirmation \cite{Walter2019,Porter2021,SwireThompson2020}. This response is not merely the result of misunderstanding; rather, it reflects deeper dynamics of identity protection, community signaling, and selective trust in sources.

The result is a form of epistemic fragmentation, where even well-designed informational interventions may have limited impact due to structural and contextual constraints \cite{swire2020searching}. Platform algorithms often prioritize content that generates engagement rather than content selected for accuracy or reliability \cite{briand2021infodemics}. At the same time, social dynamics can reinforce belief persistence through mechanisms of group identity and emotional resonance \cite{nyhan2021backfire}. In this context, misinformation may not simply reflect individual misunderstanding but can emerge as a recurrent outcome of how digital communication systems operate.

A growing body of research has begun to explore an alternative intervention strategy: prebunking \cite{van2022misinformation,basol2021towards}. Unlike debunking, which seeks to correct false beliefs after exposure, prebunking operates preemptively by exposing users to weakened or generic versions of common manipulative tactics before they encounter actual misinformation \cite{van2024inoculation}. This approach leverages principles from inoculation theory and has been shown to reduce susceptibility to false claims across domains and populations \cite{traberg2022psychological}. Unlike post-hoc corrections, which often fail due to motivational resistance or selective exposure, prebunking can act upstream—buffering cognitive vulnerabilities before belief consolidation occurs \cite{van2020psychological}. However, its effectiveness depends on timely delivery, contextual relevance, and integration into platform-level design, highlighting the need for scalable and adaptive deployment strategies.

\subsection{Coordinated Behavior and Collective Signaling}
\label{sec:coordination}

One of the most impactful dynamics in digital ecosystems is the emergence of coordinated behavior \cite{pacheco2021uncovering, Tardelli2024temporal, Cinelli2022}, often involving non-human actors or strategically organized groups \cite{hristakieva2022spread, weber2021amplifying}. Automated accounts (bots) \cite{ferrara2016rise, Cresci2020decade}, sockpuppets, and troll networks \cite{Cheng2017anyone} are frequently used to simulate authentic activity, manipulate engagement metrics, and distort perceptions of consensus \cite{Nizzoli2021,badawy2018analyzing,Shu2020,Starbird2019disinformation}. These practices—collectively referred to as coordinated inauthentic behavior (CIB)—can amplify targeted narratives, suppress alternative viewpoints, or create the illusion of spontaneous grassroots support (astroturfing) \cite{stella2018bots,luceri2019red, keller2020political, schoch2022coordination, Ratkiewicz2021detecting}. By generating high-volume, synchronized signals across multiple accounts, such campaigns can hijack visibility algorithms, flood discourse spaces, and influence user perception of legitimacy or popularity \cite{ratkiewicz2011truthy,badawy2018analyzing, Cinelli2019information}.

Alongside these artificial interventions, platforms also exhibit forms of spontaneous coordination, where large numbers of users engage in similar behaviors without centralized planning \cite{bennett2012logic,weller2013twitter}. Trends, memes, and hashtag cascades can create synchronized patterns of attention and signaling, often driven by affective contagion or identity salience rather than strategic intent \cite{castillo2014characterizing,pacheco2021uncovering}. Understanding the spectrum from scripted manipulation to emergent synchronization is key to interpreting how influence and visibility are shaped in digital environments \cite{orabi2020detection,edwards2014bot}.

Detecting these behaviors requires methodological precision \cite{alothali2018detecting, Cresci2020decade, Pant2025beyond}. Recent studies have employed similarity metrics, temporal correlation, content redundancy, and network structural features to identify coordinated entities across platforms \cite{DiMarco2025, Tardelli2024temporal, Minici2025iohunter, Nwala2023language, Cinus2025exposing}. In the context of political or toxic content, coordinated accounts tend to exhibit bursty activation patterns during high-salience events—behaving in synchrony, targeting the same content, and frequently escaping detection by moderation systems \cite{Cinelli2021dynamics, Loru2024, Cresci2019capability}. While these operations are often short-lived, they can exert outsized influence on narrative framing, agenda saturation, and emotional tone. However, their systemic impact is asymmetrical: spikes in coordinated activity coincide with strategic moments, while the background levels of toxicity or polarization in non-coordinated populations remain comparatively stable.

% Importantly, not all coordination is the product of intentional orchestration. Platforms also facilitate endogenous collective dynamics, where synchronization emerges from the interaction between attention, narrative framing, and behavioral thresholds. Trends, hashtags, memes, and symbolic cues can generate large-scale convergence without centralized planning \cite{}. In such cases, alignment is driven not by explicit design, but by shared salience signals, visibility loops, and localized amplification. These processes can elevate minority viewpoints, create the appearance of consensus, or rapidly shift the perceived agenda, often through purely organic mechanisms.
% Both forms of coordination—strategic and emergent—are mediated by platform design. 

The architecture of social systems, recommendation engines, and interaction surfaces shapes not only who sees what, but also how users align their behavior. Understanding these dynamics requires models that capture temporal coupling, feedback sensitivity, and network externalities.

Coordination represents a key point of convergence between individual behavior and systemic structure. Alongside attention dynamics, agenda setting, algorithmic amplification, and misinformation diffusion, it constitutes a foundational dimension in the study of online social systems. A cumulative science of digital behavior must account for both engineered manipulation and emergent synchronization, recognizing them as distinct but interacting drivers of visibility, perception, and influence.

\section{Modeling Opinion and Information Dynamics}
\label{sec:opinion_dynamics}

Modeling opinion dynamics is a long-standing interdisciplinary effort spanning statistical physics, computational theory, and the social sciences. These models seek to formalize how individuals form, adjust, or reinforce their beliefs through social interaction, typically under uncertainty, bounded information, and network constraints~\cite{Castellano2009,rainer2002opinion}.

Before the advent of digital platforms and large-scale behavioral datasets, these models operated largely in an abstract space. Their strength resided in analytical tractability and conceptual clarity, not in empirical correspondence~\cite{flache2017models,das2014modeling}. The increasing availability of high-resolution digital trace data has transformed this landscape~\cite{adjerid2018big}. Social media platforms now enable direct observation of belief-related behaviors—expressed through likes, shares, comments, and content exposure—across massive user populations and diverse contexts. This shift has introduced new methodological opportunities: recent work has developed procedures for learning model parameters from behavioral traces~\cite{monti2020learning}, and systematic efforts are underway to bridge the gap between formal models and real-world data~\cite{peralta2022opinion}. As a result, empirical adequacy has become a central criterion: models must not only be theoretically coherent, but also capture observable regularities in opinion and information dynamics.

Historically, the first wave of formal models drew heavily from analogies with physical systems. Weidlich’s early formulation and subsequent adaptations of the Ising model conceptualized social influence as a process of local alignment: agents tended to adopt the most common opinion in their neighborhood \cite{Weidlich1971,Galam1982,Galam1991}. These approaches revealed how global order could emerge from simple local rules, but they relied on strong assumptions—such as symmetric influence, spatial homogeneity, and memoryless dynamics—that limit their applicability in the context of digital communication. Recent reviews have highlighted these limitations and called for models that better reflect the heterogeneity and complexity of real-world settings \cite{noorazar2020classical,carpentras2023we}. Several works have integrated features such as algorithmic filtering~\cite{kertesz2019algorithmic}, personalization mechanisms~\cite{perra2019modelling}, and cognitive constraints~\cite{mei2022micro, mei2024convergence}. Others have introduced nonlinear update rules or convergence schemes that depart from naive averaging, offering a closer match to observed opinion trajectories~\cite{bizyaeva2022nonlinear}.

A second and highly influential class of models is represented by the voter model family~\cite{CLIFFORD1973,Holley1975}. In this framework, agents hold binary opinions and update their state by copying the opinion of a randomly selected neighbor. Classical extensions have explored the impact of network topology, stochastic perturbations, and bounded-confidence constraints~\cite{Redner2001,Liggett1985,DallAsta2007,Granovsky1995,Zillio2005,Antal2006,Oborny2023}, yielding key insights into metrics like consensus time, fixation probability, and noise-driven transitions.
However, their explanatory reach is limited in modern digital ecosystems, where exposure is filtered by algorithms, influence is asymmetric, and identity-based signaling matters. Recent models address these gaps by incorporating algorithmic bias into update rules and network interactions~\cite{peralta2021opinion,kertesz2019algorithmic}. 
Other work extends the voter framework by embedding personalization effects and attention decay, illustrating how polarization regimes and convergence rates shift dramatically under such mechanisms~\cite{perra2019modelling}. Empirical-oriented models advance even further: for instance, micro-foundational frameworks use weighted-median updating that align better with observed shifts in opinion formation~\cite{mei2022micro,carpentras2023we}. These developments mark a significant move toward bridging theory with behavioral data.

A third influential class is the majority‑rule model: in each round, a randomly selected group of agents updates to the majority opinion in that group. Classic results characterize convergence rates, fixation probability, and scaling behavior as functions of group size and network dimension~\cite{krapivsky2003dynamics,krapivsky2021divergence}. However, these models often rely on simplifying assumptions—static group structure, uniform mixing, and no external bias—that are misaligned with digitally mediated social systems.
Recent extensions introduce greater realism: majority-rule on hypergraphs reveals how modular or heterogeneous group connections alter convergence and prevent universal consensus~\cite{noonan2021dynamics}. Other models incorporate directed or asymmetric influence—analogous to algorithmic recommendation or propaganda—to demonstrate how slight external bias can sustain long-term polarization~\cite{forgerini2024directed,sah2024majority}. Monte-Carlo results from finite-size majority-rule systems further illustrate noise-driven transitions and the emergence of persistent coexistence regimes under realistic filtering conditions~\cite{montes2011majority}.

Along the same path, bounded confidence models (BCMs) have been introduced to capture a key empirical feature of online opinion dynamics: selective exposure. In these models, agents hold continuous opinions and interact only with others whose views lie within a defined similarity threshold~\cite{Chatterjee1977,Stone1961,Deffuant2000,rainer2002opinion,FORTUNATO2004,GMEZSERRANO2012,Piccoli2021}. The BCM framework has proven flexible in reproducing polarization, fragmentation, and stable pluralism. It accommodates the empirical observation that people tend to avoid opinion exchanges perceived as too distant or hostile.

However, most implementations rely on numerical simulations and rarely incorporate empirically derived parameters. While BCMs succeed in capturing the form of polarization, they often fail to explain the persistence and asymmetry observed in real systems—especially when echo chambers, identity-driven engagement, or asymmetric exposure are present. Recent extensions have begun addressing these issues: models that integrate heterogeneous sociability show increased fragmentation and slower convergence~\cite{li2025bounded}, while adaptive confidence bounds have been proposed to reflect individual variation in openness~\cite{li2024some}. Other works embed algorithmic influence or recommender systems into the interaction structure, amplifying polarization even under nominally moderate parameters~\cite{varshney2014bounded}. The inclusion of repulsive dynamics, where agents actively avoid disagreeing peers, has been shown to generate persistent disagreement and opinion divergence~\cite{giraldez2022analyzing}. In parallel, models that explore the emergence of consensus in networked bounded-confidence systems have highlighted the importance of structural constraints and local connectivity~\cite{javarone2014social}. Finally, multilayer models that integrate media competition and interpersonal influence—such as the framework by ~\cite{quattrociocchi2014opinion}—demonstrate how fragmented equilibria can emerge from the interplay between media messaging and social filtering in complex networks.

Despite their conceptual richness, most models in this field suffer from a structural limitation: the lack of systematic empirical validation. Often developed in synthetic environments and parameterized without reference to observed behavior, these models provide insight into possible mechanisms, but rarely offer testable predictions. This limitation is not merely technical—it reflects a deeper epistemological tension between abstract theorizing and data-driven science~\cite{adjerid2018big}. Without empirical grounding, models risk becoming internally consistent yet externally irrelevant, disconnected from the behavioral regularities that characterize real social systems. As a result, their ability to inform platform design, forecast polarization, or guide policy interventions remains limited~\cite{monti2020learning}.

Recent work has begun to address this limitation. Notably, \cite{Valensise2023} introduces and empirically validates a model of polarization using behavioral data from four major web-based social platforms. Their approach embeds empirical measures of exposure and engagement directly into the model, enabling a systematic comparison between simulated dynamics and real-world opinion trajectories. In addition, the study benchmarks its performance against alternative modeling frameworks~\cite{Baumann2020,FerrazdeArruda2022}, highlighting how different assumptions affect the ability to reproduce observed distributions. This kind of comparative, data-anchored modeling marks a necessary shift—from metaphorical abstraction toward empirically grounded explanation.
The most promising direction for the field lies in models that retain analytical clarity while being structurally constrained by empirical data—models that do not merely describe what could happen in theory, but what does happen in practice, across platforms, populations, and time.

\section{Comparative Analysis at Scale}
\label{sec:comparative}
The proliferation of cross-platform and longitudinal research has transformed the empirical baseline for modeling online social behavior. Rather than treating each platform as an isolated case, recent studies trace behavioral patterns that persist across platforms and time, providing a more reliable foundation for validating computational models.

For example, in \cite{Avalle2024,DiMarco2024}) the authors document phenomena such as increasing toxicity in longer thread conversations and systematic simplification of user language across platforms and time. 

These results support a critical insight: algorithmic systems may not create conversational dysfunction—they often expose and amplify human behavior that is consistent across platforms. This insight has important implications for model builders: theories of opinion dynamics should be tested not only for algorithmic effects but also for their capacity to capture structural regularities intrinsic to large-scale mediated interaction.

Crucially, the multiplatform strategy enables researchers to distinguish between system-specific artifacts and structural regularities \cite{singh2024differences,baqir2024news}. It allows for the formulation of more robust hypotheses about causal mechanisms, enabling inference that transcends individual platform biases \cite{saberski2024impact,murdock2024information}. The integration of large-scale trace data, computational modeling, and comparative analysis opens the door to a new class of hybrid models: empirically informed, mechanistically explicit, and cross-context validated \cite{Avalle2024}.

What emerges from this literature is a clear imperative: no model, however elegant, parsimonious, or theoretically compelling, can be considered explanatory unless it is validated against observed behavioral patterns \cite{wagner2021measuring}. The need is especially pressing in algorithmically infused societies, where structural dynamics must be rigorously measured based on empirical data. Recent cross-platform analyses uncover invariant conversational and linguistic patterns—such as persistent toxicity dynamics and systematic simplification of language—that transcend platform-specific affordances \cite{Avalle2024,DiMarco2024}.

Future models must internalize these dimensions—either by embedding them structurally or by calibrating parameters on real-world data. Without such alignment, models risk remaining metaphorical abstractions, disconnected from the dynamics they aim to explain.

In sum, the study of opinion dynamics stands at a critical juncture. Decades of formal modeling have yielded elegant abstractions, but abstraction alone no longer suffices. The rise of digital platforms has introduced a structural discontinuity: opinion formation now unfolds within socio-technical systems that generate massive, traceable, and algorithmically shaped behavioral data. This transformation demands more than theoretical refinement—it calls for empirical accountability. Only by confronting this dual imperative—preserving conceptual clarity while anchoring models in observed regularities—can the field evolve from metaphor to mechanism. Without this convergence, models will remain speculative blueprints. With it, they can become diagnostic instruments for decoding and ultimately counteracting the systemic pathologies of online collective behavior.

\section{Limitations}
\label{sec:limitations}

Over the past decade, research on online behavior has revealed a growing number of empirical regularities: power-law distributions of attention, echo chambers, asymmetric exposure, and recurrent patterns of coordination. These findings, often replicated across platforms and contexts, represent an important step toward a more systematic understanding of the dynamics underlying digital interactions. Yet despite this empirical progress, the field still lacks a shared analytical framework to explain how such patterns emerge, under what conditions they persist, and how they interact with platform design.

The first open challenge concerns the development of models that capture behavioral dynamics beyond idealized scenarios. Many current models remain loosely connected to empirical observables and rely on assumptions that are difficult to validate or calibrate with real-world data. More systematic efforts are needed to link model parameters to measurable quantities and to evaluate explanatory power beyond internal consistency or stylized reproduction.

A second limitation lies in the difficulty of isolating the role of platform architecture. While similar behavioral phenomena appear across diverse systems, the contribution of algorithmic curation, moderation policies, and interface design remains difficult to disentangle from endogenous dynamics. Comparative studies, leveraging natural variation across platforms or temporal policy shifts, may offer a viable strategy, but require coordinated efforts in data collection and methodological design.

Causal inference poses a third conceptual and practical challenge. Randomized experiments, though powerful, are rare and typically limited to specific partnerships or restricted user groups. Alternative strategies—such as quasi-experimental designs, instrumental variables, or simulation-informed counterfactuals—remain underused in this context. Their systematic integration into the study of digital systems could substantially improve our ability to test hypotheses about mechanism and effect.

Another underdeveloped area is the identification of observables that capture system-level dynamics over time. Much of the current literature relies on static metrics, which offer limited insight into how fragmentation, synchronization, or volatility evolve. Temporal measures—such as entropy in interaction networks, burst alignment, or shifts in exposure diversity—could offer more sensitive indicators of systemic transitions or emerging risks.

Finally, most behavioral models treat exposure as exogenous, neglecting the recursive interplay between user activity and algorithmic filtering. This simplification limits our ability to understand feedback effects, such as amplification or convergence, that may be critical in shaping outcomes. Even minimal models that incorporate adaptive exposure mechanisms could provide valuable insight into how interventions propagate through user–system coupling.

These limitations do not undermine the progress made so far, but they highlight the need for a more integrated approach, one that connects empirical findings, theoretical models, and system-level inference within a coherent analytical structure. Rather than proposing a unified agenda, we suggest that addressing these issues represents a necessary step toward a cumulative science of online behavior, capable of moving from descriptive mapping to explanatory understanding.
Clarifying these foundational aspects may also help inform future system design and policy evaluation, grounding interventions in empirically supported mechanisms.

\section{Design Implications}
\label{sec:design}

The behavioral patterns identified across platforms — such as concentration of attention, ideological clustering, and content volatility — are not solely the product of user preferences. They often emerge from the interaction between individual behavior and system-level affordances, including ranking algorithms, interface structures, and feedback mechanisms. Understanding these dynamics offers concrete directions for system design.
Despite the limitations discussed in the previous sections, these scientific findings can now serve as a foundation for identifying key implications and informing future developments.

In this view, we outline below a set of design-relevant implications that follow from the empirical and modeling literature:

\begin{itemize}
  \item \textbf{Real-time indicators of systemic change.} Many platform diagnostics still focus primarily on micro-level engagement metrics and rely on data accessible only to the platforms themselves. However, providing researchers and moderators with access to longitudinal data and measures could enable earlier detection of patterns such as fragmentation, synchronization, or volatility.

  \item \textbf{Balance engagement signals with structural considerations.} Algorithmic curation often optimizes for short-term engagement, which can amplify emotionally charged or polarizing content. Incorporating complementary signals such as diversity of exposure, novelty, or reputational indicators may help counteract these tendencies without disrupting relevance.

  \item \textbf{Introduce calibrated exposure diversity.} Personalization systems tend to reinforce existing preferences, narrowing informational input. Design strategies that support controlled variation may preserve user relevance while increasing viewpoint plurality.

  \item \textbf{Enhance interpretability of content ranking.} Users rarely understand why specific content appears in their feed. Simple, human-readable explanations of ranking decisions could improve transparency and support external auditing efforts, without requiring full algorithmic disclosure.

  \item \textbf{Incorporate feedback sensitivity into design evaluation.} User behavior and algorithmic filtering interact recursively. Even minimal models that include adaptive feedback loops can help anticipate amplification effects or behavioral convergence. Integrating such models into system evaluation pipelines may improve robustness to unintended outcomes.
\end{itemize}

These implications do not prescribe normative goals for platform governance. Rather, they highlight structural levers through which visibility, diversity, and stability can be modulated. Embedding behavioral insights into design processes represents a critical step toward systems that are not only efficient but also resilient to manipulation, fragmentation, and epistemic degradation.
Moreover, it is important to emphasize that these changes, when paired with appropriate data access, will enable researchers to better understand the mechanisms underlying online behavior.

\section{Conclusions}
\label{sec:conclusion}
The emergence of digital platforms has redefined the structural conditions under which information is produced, disseminated, and internalized. This transformation is not transient; it marks a long-term reconfiguration of the informational landscape, with implications for how social behavior is observed, modeled, and interpreted. The convergence of algorithmic mediation, user-generated content, and large-scale behavioral traces has enabled a shift in the study of online dynamics—from anecdotal description to systematic empirical observation.

This survey has synthesized findings across key domains. In each, computational approaches have uncovered statistical regularities that challenge long-standing assumptions and suggest the existence of structural patterns that transcend specific platforms or contexts. Phenomena such as echo chambers, misinformation cascades, linguistic simplification, and polarization no longer appear as isolated effects but as outcomes shaped by the interplay between user behavior, algorithmic logic, and socio-technical affordances.

Yet despite substantial progress, the field continues to face persistent challenges, as highlighted in Section \ref{sec:limitations}.

% First, the methodological gap between theoretical models and empirical observations limits explanatory power. Many models remain under-validated, and their assumptions are rarely tested against behavioral data at scale. The availability of large-scale, multiplatform datasets opens new opportunities for empirical calibration, but also raises the bar for what constitutes an adequate model of social dynamics.

% Second, the infrastructural conditions for behavioral research remain fragile. Access to meaningful platform data is increasingly restricted, fragmented across closed APIs, or governed by opaque gatekeeping practices. These limitations hinder replicability, constrain the scope of inquiry, and create dependencies that compromise scientific independence. Addressing this requires not only technical solutions but new institutional frameworks that support open, secure, and ethically governed access to behavioral data.

% Third, much empirical work remains confined to individual platforms, limiting the generalizability of results. Single-platform studies risk mistaking local artifacts for global dynamics. Recent longitudinal and cross-platform research suggests that some behavioral patterns—such as the escalation of toxicity in long discussions or the simplification of user language—are robust across systems. Moving from platform-specific studies to a comparative science of digital behavior is essential for identifying structural invariants and informing the design of reliable models.

Addressing them is necessary for the development of a cumulative research program in online social dynamics. It calls for coordinated efforts to establish validated pipelines, interoperable observatories, and public infrastructures for behavioral data access. Most critically, it requires a shared commitment to methodological rigor and empirical accountability across the research community.

The promise of computational social science lies in its ability to render complex behavioral systems observable, interpretable, and, ultimately, actionable. Yet the term itself increasingly fails to capture the epistemic shift underway. Rather than a stable disciplinary identity, computational social science has often functioned as an umbrella category—broad in scope but weak in methodological cohesion. What is now emerging is a more integrated paradigm: one grounded in the analytical tools of data science, the formal frameworks of complex systems, and the empirical richness of large-scale social data. This convergence marks a transition—from a loose interdiscipline to a structured scientific approach capable of explanatory depth and predictive precision.

Future work should proceed along four converging directions. First, the development of empirically validated models that integrate cognitive constraints, algorithmic mediation, and observed behavioral patterns across platforms. Second, the construction of multiplatform observatories that support synchronized, longitudinal data collection under consistent ethical and technical protocols. Third, the advancement of causal inference techniques—such as counterfactual simulations, synthetic controls, and large-scale interventions—that can disentangle behavioral dynamics from infrastructural effects. Fourth, the institutionalization of open, independent infrastructures for behavioral data access, designed to ensure reproducibility, transparency, and long-term availability of high-quality digital trace data. These directions are not incremental refinements; they define the structural foundations for a robust and cumulative science of online behavior.

% \appendix
% \section{Example Appendix Section}





\section*{Acknowledgement}
The work is supported by IRIS Infodemic Coalition (UK government, grant no. SCH-00001-3391), 
SERICS (PE00000014) under the NRRP MUR program funded by the European Union - NextGenerationEU, project CRESP from the Italian Ministry of Health under the program CCM 2022, PON project “Ricerca e Innovazione” 2014-2020 and project SEED n. SP122184858BEDB3.

We would also like to thank the Hypnotoad for inspiring our research and discussion. 

\bibliographystyle{unsrt}
\bibliography{bibliography}

\begin{thebibliography}{100}

\bibitem{del2016spreading}
Michela Del~Vicario, Alessandro Bessi, Fabiana Zollo, Fabio Petroni, Antonio Scala, Guido Caldarelli, H~Eugene Stanley, and Walter Quattrociocchi.
\newblock The spreading of misinformation online.
\newblock {\em Proceedings of the national academy of Sciences}, 113(3):554--559, 2016.

\bibitem{vosoughi2018spread}
Soroush Vosoughi, Deb Roy, and Sinan Aral.
\newblock The spread of true and false news online.
\newblock {\em science}, 359(6380):1146--1151, 2018.

\bibitem{schmidt2017anatomy}
Ana~Luc{\'\i}a Schmidt, Fabiana Zollo, Michela Del~Vicario, Alessandro Bessi, Antonio Scala, Guido Caldarelli, H~Eugene Stanley, and Walter Quattrociocchi.
\newblock Anatomy of news consumption on facebook.
\newblock {\em Proceedings of the National Academy of Sciences}, 114(12):3035--3039, 2017.

\bibitem{kumpel2015news}
Anna~Sophie K{\"u}mpel, Veronika Karnowski, and Till Keyling.
\newblock News sharing in social media: A review of current research on news sharing users, content, and networks.
\newblock {\em Social media+ society}, 1(2):2056305115610141, 2015.

\bibitem{tsagkias2011linking}
Manos Tsagkias, Maarten De~Rijke, and Wouter Weerkamp.
\newblock Linking online news and social media.
\newblock In {\em Proceedings of the fourth ACM international conference on Web search and data mining}, pages 565--574, 2011.

\bibitem{centola2010spread}
Damon Centola.
\newblock The spread of behavior in an online social network experiment.
\newblock {\em science}, 329(5996):1194--1197, 2010.

\bibitem{johnson2019hidden}
Neil~F Johnson, Rhys Leahy, N~Johnson Restrepo, Nicholas Vel{\'a}squez, Minzhang Zheng, Pedro Manrique, Prajwal Devkota, and Stefan Wuchty.
\newblock Hidden resilience and adaptive dynamics of the global online hate ecology.
\newblock {\em Nature}, 573(7773):261--265, 2019.

\bibitem{guy2010social}
Ido Guy, Naama Zwerdling, Inbal Ronen, David Carmel, and Erel Uziel.
\newblock Social media recommendation based on people and tags.
\newblock In {\em Proceedings of the 33rd international ACM SIGIR conference on Research and development in information retrieval}, pages 194--201, 2010.

\bibitem{momeni2015survey}
Elaheh Momeni, Claire Cardie, and Nicholas Diakopoulos.
\newblock A survey on assessment and ranking methodologies for user-generated content on the web.
\newblock {\em ACM Computing Surveys (CSUR)}, 48(3):1--49, 2015.

\bibitem{gillespie2018custodians}
Tarleton Gillespie.
\newblock {\em Custodians of the Internet: Platforms, content moderation, and the hidden decisions that shape social media}.
\newblock Yale University Press, 2018.

\bibitem{Bail2018}
Christopher~A. Bail, Lisa~P. Argyle, Taylor~W. Brown, John~P. Bumpus, Haohan Chen, M.~B.~Fallin Hunzaker, Jaemin Lee, Marcus Mann, Friedolin Merhout, and Alexander Volfovsky.
\newblock Exposure to opposing views on social media can increase political polarization.
\newblock {\em Proceedings of the National Academy of Sciences}, 115(37):9216–9221, August 2018.

\bibitem{tucker2018social}
Joshua~A Tucker, Andrew Guess, Pablo Barber{\'a}, Cristian Vaccari, Alexandra Siegel, Sergey Sanovich, Denis Stukal, and Brendan Nyhan.
\newblock Social media, political polarization, and political disinformation: A review of the scientific literature.
\newblock {\em Political polarization, and political disinformation: a review of the scientific literature (March 19, 2018)}, 2018.

\bibitem{injadat2016data}
MohammadNoor Injadat, Fadi Salo, and Ali~Bou Nassif.
\newblock Data mining techniques in social media: A survey.
\newblock {\em Neurocomputing}, 214:654--670, 2016.

\bibitem{flaxman2016filter}
Seth Flaxman, Sharad Goel, and Justin~M Rao.
\newblock Filter bubbles, echo chambers, and online news consumption.
\newblock {\em Public opinion quarterly}, 80(S1):298--320, 2016.

\bibitem{guess2018selective}
Andrew Guess, Brendan Nyhan, and Jason Reifler.
\newblock Selective exposure to misinformation: Evidence from the consumption of fake news during the 2016 us presidential campaign.
\newblock {\em European Research Council}, 9(3):4, 2018.

\bibitem{budak2024misunderstanding}
Ceren Budak, Brendan Nyhan, David~M Rothschild, Emily Thorson, and Duncan~J Watts.
\newblock Misunderstanding the harms of online misinformation.
\newblock {\em Nature}, 630(8015):45--53, 2024.

\bibitem{bessi2015science}
Alessandro Bessi, Mauro Coletto, George~Alexandru Davidescu, Antonio Scala, Guido Caldarelli, and Walter Quattrociocchi.
\newblock Science vs conspiracy: Collective narratives in the age of misinformation.
\newblock {\em PloS one}, 10(2):e0118093, 2015.

\bibitem{cinelli2021echo}
Matteo Cinelli, Gianmarco De~Francisci~Morales, Alessandro Galeazzi, Walter Quattrociocchi, and Michele Starnini.
\newblock The echo chamber effect on social media.
\newblock {\em Proceedings of the national academy of sciences}, 118(9):e2023301118, 2021.

\bibitem{garimella2018political}
Kiran Garimella, Gianmarco De~Francisci~Morales, Aristides Gionis, and Michael Mathioudakis.
\newblock Political discourse on social media: Echo chambers, gatekeepers, and the price of bipartisanship.
\newblock In {\em Proceedings of the 2018 world wide web conference}, pages 913--922, 2018.

\bibitem{barbera2015tweeting}
Pablo Barber{\'a}, John~T Jost, Jonathan Nagler, Joshua~A Tucker, and Richard Bonneau.
\newblock Tweeting from left to right: Is online political communication more than an echo chamber?
\newblock {\em Psychological science}, 26(10):1531--1542, 2015.

\bibitem{del2017mapping}
Michela Del~Vicario, Fabiana Zollo, Guido Caldarelli, Antonio Scala, and Walter Quattrociocchi.
\newblock Mapping social dynamics on facebook: The brexit debate.
\newblock {\em Social Networks}, 50:6--16, 2017.

\bibitem{cinelli2020covid}
Matteo Cinelli, Walter Quattrociocchi, Alessandro Galeazzi, Carlo~Michele Valensise, Emanuele Brugnoli, Ana~Lucia Schmidt, Paola Zola, Fabiana Zollo, and Antonio Scala.
\newblock The covid-19 social media infodemic.
\newblock {\em Scientific reports}, 10(1):16598, 2020.

\bibitem{do2022infodemics}
Israel Junior~Borges Do~Nascimento, Ana~Beatriz Pizarro, Jussara~M Almeida, Natasha Azzopardi-Muscat, Marcos~Andr{\'e} Gon{\c{c}}alves, Maria Bj{\"o}rklund, and David Novillo-Ortiz.
\newblock Infodemics and health misinformation: a systematic review of reviews.
\newblock {\em Bulletin of the World Health Organization}, 100(9):544, 2022.

\bibitem{alvarez2021determinants}
Javier Alvarez-Galvez, Victor Suarez-Lledo, and Antonio Rojas-Garcia.
\newblock Determinants of infodemics during disease outbreaks: a systematic review.
\newblock {\em Frontiers in public health}, 9:603603, 2021.

\bibitem{bashir2017effects}
Hilal Bashir and Shabir~Ahmad Bhat.
\newblock Effects of social media on mental health: A review.
\newblock {\em International Journal of Indian Psychology}, 4(3):125--131, 2017.

\bibitem{pierri2022online}
Francesco Pierri, Brea~L Perry, Matthew~R DeVerna, Kai-Cheng Yang, Alessandro Flammini, Filippo Menczer, and John Bryden.
\newblock Online misinformation is linked to early covid-19 vaccination hesitancy and refusal.
\newblock {\em Scientific reports}, 12(1):5966, 2022.

\bibitem{burki2019vaccine}
Talha Burki.
\newblock Vaccine misinformation and social media.
\newblock {\em The Lancet Digital Health}, 1(6):e258--e259, 2019.

\bibitem{Orben2022}
Amy Orben, Andrew~K. Przybylski, Sarah-Jayne Blakemore, and Rogier~A. Kievit.
\newblock Windows of developmental sensitivity to social media.
\newblock {\em Nature Communications}, 13(1), March 2022.

\bibitem{Naslund2016}
J.~A. Naslund, K.~A. Aschbrenner, L.~A. Marsch, and S.~J. Bartels.
\newblock The future of mental health care: peer-to-peer support and social media.
\newblock {\em Epidemiology and Psychiatric Sciences}, 25(2):113–122, January 2016.

\bibitem{weigle2024social}
Paul~E Weigle and Reem~MA Shafi.
\newblock Social media and youth mental health.
\newblock {\em Current psychiatry reports}, 26(1):1--8, 2024.

\bibitem{ferguson2024there}
Christopher~J Ferguson, Linda~K Kaye, Dawn Branley-Bell, and Patrick Markey.
\newblock There is no evidence that time spent on social media is correlated with adolescent mental health problems: Findings from a meta-analysis.
\newblock {\em Professional Psychology: Research and Practice}, 2024.

\bibitem{pecile2025mapping}
Giulio Pecile, Niccol{\`o} Di~Marco, Matteo Cinelli, and Walter Quattrociocchi.
\newblock Mapping the global election landscape on social media in 2024.
\newblock {\em PloS one}, 20(2):e0316271, 2025.

\bibitem{zhuravskaya2020political}
Ekaterina Zhuravskaya, Maria Petrova, and Ruben Enikolopov.
\newblock Political effects of the internet and social media.
\newblock {\em Annual review of economics}, 12(1):415--438, 2020.

\bibitem{metaxas2012social}
Panagiotis~T Metaxas and Eni Mustafaraj.
\newblock Social media and the elections.
\newblock {\em Science}, 338(6106):472--473, 2012.

\bibitem{aral2019protecting}
Sinan Aral and Dean Eckles.
\newblock Protecting elections from social media manipulation.
\newblock {\em Science}, 365(6456):858--861, 2019.

\bibitem{ruths2014social}
Derek Ruths and J{\"u}rgen Pfeffer.
\newblock Social media for large studies of behavior.
\newblock {\em Science}, 346(6213):1063--1064, 2014.

\bibitem{lazer2020computational}
David~MJ Lazer, Alex Pentland, Duncan~J Watts, Sinan Aral, Susan Athey, Noshir Contractor, Deen Freelon, Sandra Gonzalez-Bailon, Gary King, Helen Margetts, et~al.
\newblock Computational social science: Obstacles and opportunities.
\newblock {\em Science}, 369(6507):1060--1062, 2020.

\bibitem{gonzalez2011dynamics}
Sandra Gonz{\'a}lez-Bail{\'o}n, Javier Borge-Holthoefer, Alejandro Rivero, and Yamir Moreno.
\newblock The dynamics of protest recruitment through an online network.
\newblock {\em Scientific reports}, 1(1):1--7, 2011.

\bibitem{Avalle2024}
Michele Avalle, Niccolò Di~Marco, Gabriele Etta, Emanuele Sangiorgio, Shayan Alipour, Anita Bonetti, Lorenzo Alvisi, Antonio Scala, Andrea Baronchelli, Matteo Cinelli, and Walter Quattrociocchi.
\newblock Persistent interaction patterns across social media platforms and over time.
\newblock {\em Nature}, 628(8008):582–589, March 2024.

\bibitem{wagner2021measuring}
Claudia Wagner, Markus Strohmaier, Alexandra Olteanu, Emre K{\i}c{\i}man, Noshir Contractor, and Tina Eliassi-Rad.
\newblock Measuring algorithmically infused societies.
\newblock {\em Nature}, 595(7866):197--204, 2021.

\bibitem{lazer2009computational}
David Lazer, Alex Pentland, Lada Adamic, Sinan Aral, Albert-L{\'a}szl{\'o} Barab{\'a}si, Devon Brewer, Nicholas Christakis, Noshir Contractor, James Fowler, Myron Gutmann, et~al.
\newblock Computational social science.
\newblock {\em Science}, 323(5915):721--723, 2009.

\bibitem{Conte2012}
R.~Conte, N.~Gilbert, G.~Bonelli, C.~Cioffi-Revilla, G.~Deffuant, J.~Kertesz, V.~Loreto, S.~Moat, J.~P. Nadal, A.~Sanchez, A.~Nowak, A.~Flache, M.~San~Miguel, and D.~Helbing.
\newblock Manifesto of computational social science.
\newblock {\em The European Physical Journal Special Topics}, 214(1):325–346, November 2012.

\bibitem{watts2007twenty}
Duncan~J Watts.
\newblock A twenty-first century science.
\newblock {\em Nature}, 445(7127):489--489, 2007.

\bibitem{gonzalez2016networked}
Sandra Gonz{\'a}lez-Bail{\'o}n and Ning Wang.
\newblock Networked discontent: The anatomy of protest campaigns in social media.
\newblock {\em Social networks}, 44:95--104, 2016.

\bibitem{mosleh2020self}
Mohsen Mosleh, Gordon Pennycook, and David~G Rand.
\newblock Self-reported willingness to share political news articles in online surveys correlates with actual sharing on twitter.
\newblock {\em Plos one}, 15(2):e0228882, 2020.

\bibitem{tufekci2014big}
Zeynep Tufekci.
\newblock Big questions for social media big data: Representativeness, validity and other methodological pitfalls.
\newblock In {\em Proceedings of the international AAAI conference on web and social media}, volume~8, pages 505--514, 2014.

\bibitem{wang2023less}
Xiaohui Wang, Yunya Song, and Youzhen Su.
\newblock Less fragmented but highly centralized: a bibliometric analysis of research in computational social science.
\newblock {\em Social Science Computer Review}, 41(3):946--966, 2023.

\bibitem{paakkonen2024emergence}
Juho P{\"a}{\"a}kk{\"o}nen, Matti Nelimarkka, and Samuli Reijula.
\newblock The emergence of computational social science: Intellectual integration or persistent fragmentation?
\newblock 2024.

\bibitem{pablo2014social}
Barber{\'a} Pablo.
\newblock How social media reduces mass political polarization. evidence from germany, spain, and the us.
\newblock {\em Job Market Paper}, 2014.

\bibitem{lelkes2017hostile}
Yphtach Lelkes, Gaurav Sood, and Shanto Iyengar.
\newblock The hostile audience: The effect of access to broadband internet on partisan affect.
\newblock {\em American Journal of Political Science}, 61(1):5--20, 2017.

\bibitem{vicario2019polarization}
Michela~Del Vicario, Walter Quattrociocchi, Antonio Scala, and Fabiana Zollo.
\newblock Polarization and fake news: Early warning of potential misinformation targets.
\newblock {\em ACM Transactions on the Web (TWEB)}, 13(2):1--22, 2019.

\bibitem{freelon2018computational}
Deen Freelon.
\newblock Computational research in the post-api age.
\newblock {\em Political Communication}, 35(4):665--668, 2018.

\bibitem{guess2020misinformation}
Andrew~M Guess and Benjamin~A Lyons.
\newblock Misinformation, disinformation, and online propaganda.
\newblock {\em Social media and democracy: The state of the field, prospects for reform}, 10:10--33, 2020.

\bibitem{DiMarco2024}
Niccolò Di~Marco, Edoardo Loru, Anita Bonetti, Alessandra Olga~Grazia Serra, Matteo Cinelli, and Walter Quattrociocchi.
\newblock Patterns of linguistic simplification on social media platforms over time.
\newblock {\em Proceedings of the National Academy of Sciences}, 121(50), December 2024.

\bibitem{carpenter2011evaluating}
John Carpenter.
\newblock Evaluating social work education: A review of outcomes, measures, research designs and practicalities.
\newblock {\em Social Work Education}, 30(02):122--140, 2011.

\bibitem{reid2004some}
William~J Reid, Bonnie~Davis Kenaley, and Julanne Colvin.
\newblock Do some interventions work better than others? a review of comparative social work experiments.
\newblock {\em Social work research}, 28(2):71--81, 2004.

\bibitem{tsui1997empirical}
Ming-Sum Tsui.
\newblock Empirical research on social work supervision: The state of the art (1970-1995).
\newblock {\em Journal of Social Service Research}, 23(2):39--54, 1997.

\bibitem{bell2012obstacles}
Stephen~H Bell and Laura~R Peck.
\newblock Obstacles to and limitations of social experiments: 15 false alarms.
\newblock {\em Abt thought leadership paper, Abt Associates}, 2012.

\bibitem{Button2013}
Katherine~S. Button, John P.~A. Ioannidis, Claire Mokrysz, Brian~A. Nosek, Jonathan Flint, Emma S.~J. Robinson, and Marcus~R. Munafò.
\newblock Power failure: why small sample size undermines the reliability of neuroscience.
\newblock {\em Nature Reviews Neuroscience}, 14(5):365–376, April 2013.

\bibitem{Makel2012}
Matthew~C. Makel, Jonathan~A. Plucker, and Boyd Hegarty.
\newblock Replications in psychology research: How often do they really occur?
\newblock {\em Perspectives on Psychological Science}, 7(6):537–542, November 2012.

\bibitem{Furnham1986}
Adrian Furnham.
\newblock Response bias, social desirability and dissimulation.
\newblock {\em Personality and Individual Differences}, 7(3):385–400, January 1986.

\bibitem{Rosenman2011}
Robert Rosenman, Vidhura Tennekoon, and Laura~G. Hill.
\newblock Measuring bias in self-reported data.
\newblock {\em International Journal of Behavioural and Healthcare Research}, 2(4):320, 2011.

\bibitem{Koller2023}
Katharina Koller, Paulina~K. Pankowska, and Cameron Brick.
\newblock Identifying bias in self-reported pro-environmental behavior.
\newblock {\em Current Research in Ecological and Social Psychology}, 4:100087, 2023.

\bibitem{wu2021platform}
Angela~Xiao Wu and Harsh Taneja.
\newblock Platform enclosure of human behavior and its measurement: Using behavioral trace data against platform episteme.
\newblock {\em New Media \& Society}, 23(9):2650--2667, 2021.

\bibitem{kim2017like}
Cheonsoo Kim and Sung-Un Yang.
\newblock Like, comment, and share on facebook: How each behavior differs from the other.
\newblock {\em Public relations review}, 43(2):441--449, 2017.

\bibitem{schreiner2021impact}
Melanie Schreiner, Thomas Fischer, and Rene Riedl.
\newblock Impact of content characteristics and emotion on behavioral engagement in social media: literature review and research agenda.
\newblock {\em Electronic Commerce Research}, 21:329--345, 2021.

\bibitem{Bell2009}
Gordon Bell, Tony Hey, and Alex Szalay.
\newblock Beyond the data deluge.
\newblock {\em Science}, 323(5919):1297–1298, March 2009.

\bibitem{Parry2021}
Douglas~A. Parry, Brittany~I. Davidson, Craig J.~R. Sewall, Jacob~T. Fisher, Hannah Mieczkowski, and Daniel~S. Quintana.
\newblock A systematic review and meta-analysis of discrepancies between logged and self-reported digital media use.
\newblock {\em Nature Human Behaviour}, 5(11):1535–1547, May 2021.

\bibitem{sultan2023leaving}
Mubashir Sultan, Christin Scholz, and Wouter van~den Bos.
\newblock Leaving traces behind: Using social media digital trace data to study adolescent wellbeing.
\newblock {\em Computers in Human Behavior Reports}, 10:100281, 2023.

\bibitem{kumar2011understanding}
Shamanth Kumar, Reza Zafarani, and Huan Liu.
\newblock Understanding user migration patterns in social media.
\newblock In {\em Proceedings of the AAAI Conference on Artificial Intelligence}, volume~25, pages 1204--1209, 2011.

\bibitem{adedoyin2014survey}
Mariam Adedoyin-Olowe, Mohamed~Medhat Gaber, and Frederic Stahl.
\newblock A survey of data mining techniques for social media analysis.
\newblock {\em Journal of Data Mining \& Digital Humanities}, 2014, 2014.

\bibitem{li2017survey}
Mei Li, Xiang Wang, Kai Gao, and Shanshan Zhang.
\newblock A survey on information diffusion in online social networks: Models and methods.
\newblock {\em Information}, 8(4):118, 2017.

\bibitem{guille2013information}
Adrien Guille, Hakim Hacid, Cecile Favre, and Djamel~A Zighed.
\newblock Information diffusion in online social networks: A survey.
\newblock {\em ACM Sigmod Record}, 42(2):17--28, 2013.

\bibitem{backstrom2006group}
Lars Backstrom, Dan Huttenlocher, Jon Kleinberg, and Xiangyang Lan.
\newblock Group formation in large social networks: membership, growth, and evolution.
\newblock In {\em Proceedings of the 12th ACM SIGKDD international conference on Knowledge discovery and data mining}, pages 44--54, 2006.

\bibitem{del2016echo}
Michela Del~Vicario, Gianna Vivaldo, Alessandro Bessi, Fabiana Zollo, Antonio Scala, Guido Caldarelli, and Walter Quattrociocchi.
\newblock Echo chambers: Emotional contagion and group polarization on facebook.
\newblock {\em Scientific reports}, 6(1):37825, 2016.

\bibitem{lim2022opinion}
Soo~Ling Lim and Peter~J Bentley.
\newblock Opinion amplification causes extreme polarization in social networks.
\newblock {\em Scientific Reports}, 12(1):18131, 2022.

\bibitem{huszar2022algorithmic}
Ferenc Husz{\'a}r, Sofia~Ira Ktena, Conor O’Brien, Luca Belli, Andrew Schlaikjer, and Moritz Hardt.
\newblock Algorithmic amplification of politics on twitter.
\newblock {\em Proceedings of the national academy of sciences}, 119(1):e2025334119, 2022.

\bibitem{acker2020social}
Amelia Acker and Adam Kreisberg.
\newblock Social media data archives in an api-driven world.
\newblock {\em Archival Science}, 20(2):105--123, 2020.

\bibitem{abrams2025neutrality}
Kyra~Milan Abrams and Madelyn~Rose Sanfilippo.
\newblock Neutrality or contextuality.
\newblock {\em Social Informatics}, 2025.

\bibitem{hallinan2022beyond}
Blake Hallinan, Rebecca Scharlach, and Limor Shifman.
\newblock Beyond neutrality: Conceptualizing platform values.
\newblock {\em Communication Theory}, 32(2):201--222, 2022.

\bibitem{zafarani2014behavior}
Reza Zafarani and Huan Liu.
\newblock Behavior analysis in social media.
\newblock {\em IEEE Intelligent systems}, 29(4):1--4, 2014.

\bibitem{cohen2013classifying}
Raviv Cohen and Derek Ruths.
\newblock Classifying political orientation on twitter: It’s not easy!
\newblock In {\em Proceedings of the International AAAI Conference on Web and Social Media}, volume~7, pages 91--99, 2013.

\bibitem{jurgens2015geolocation}
David Jurgens, Tyler Finethy, James McCorriston, Yi~Xu, and Derek Ruths.
\newblock Geolocation prediction in twitter using social networks: A critical analysis and review of current practice.
\newblock In {\em Proceedings of the international AAAI conference on web and social media}, volume~9, pages 188--197, 2015.

\bibitem{bovet2019influence}
Alexandre Bovet and Hern{\'a}n~A Makse.
\newblock Influence of fake news in twitter during the 2016 us presidential election.
\newblock {\em Nature communications}, 10(1):7, 2019.

\bibitem{flamino2023political}
James Flamino, Alessandro Galeazzi, Stuart Feldman, Michael~W Macy, Brendan Cross, Zhenkun Zhou, Matteo Serafino, Alexandre Bovet, Hern{\'a}n~A Makse, and Boleslaw~K Szymanski.
\newblock Political polarization of news media and influencers on twitter in the 2016 and 2020 us presidential elections.
\newblock {\em Nature Human Behaviour}, 7(6):904--916, 2023.

\bibitem{weng2013virality}
Lilian Weng, Filippo Menczer, and Yong-Yeol Ahn.
\newblock Virality prediction and community structure in social networks.
\newblock {\em Scientific reports}, 3(1):2522, 2013.

\bibitem{johnson2020online}
Neil~F Johnson, Nicolas Vel{\'a}squez, Nicholas~Johnson Restrepo, Rhys Leahy, Nicholas Gabriel, Sara El~Oud, Minzhang Zheng, Pedro Manrique, Stefan Wuchty, and Yonatan Lupu.
\newblock The online competition between pro-and anti-vaccination views.
\newblock {\em Nature}, 582(7811):230--233, 2020.

\bibitem{park2003hyperlink}
Han~Woo Park and Mike Thelwall.
\newblock Hyperlink analyses of the world wide web: A review.
\newblock {\em Journal of computer-mediated communication}, 8(4):JCMC843, 2003.

\bibitem{bandy2023facebook}
Jack Bandy and Nicholas Diakopoulos.
\newblock Facebook’s news feed algorithm and the 2020 us election.
\newblock {\em Social Media+ Society}, 9(3):20563051231196898, 2023.

\bibitem{oeldorf2020ineffectiveness}
Anne Oeldorf-Hirsch, Mike Schmierbach, Alyssa Appelman, and Michael~P Boyle.
\newblock The ineffectiveness of fact-checking labels on news memes and articles.
\newblock {\em Mass Communication and Society}, 23(5):682--704, 2020.

\bibitem{martel2024fact}
Cameron Martel and David~G Rand.
\newblock Fact-checker warning labels are effective even for those who distrust fact-checkers.
\newblock {\em Nature Human Behaviour}, 8(10):1957--1967, 2024.

\bibitem{lazer2015rise}
David Lazer.
\newblock The rise of the social algorithm.
\newblock {\em Science}, 348(6239):1090--1091, 2015.

\bibitem{lewandowsky2021countering}
Stephan Lewandowsky and Sander Van Der~Linden.
\newblock Countering misinformation and fake news through inoculation and prebunking.
\newblock {\em European review of social psychology}, 32(2):348--384, 2021.

\bibitem{mosleh2022field}
Mohsen Mosleh, Gordon Pennycook, and David~G Rand.
\newblock Field experiments on social media.
\newblock {\em Current Directions in Psychological Science}, 31(1):69--75, 2022.

\bibitem{bartley2021auditing}
Nathan Bartley, Andres Abeliuk, Emilio Ferrara, and Kristina Lerman.
\newblock Auditing algorithmic bias on twitter.
\newblock In {\em Proceedings of the 13th ACM Web Science Conference 2021}, pages 65--73, 2021.

\bibitem{shen2021everyday}
Hong Shen, Alicia DeVos, Motahhare Eslami, and Kenneth Holstein.
\newblock Everyday algorithm auditing: Understanding the power of everyday users in surfacing harmful algorithmic behaviors.
\newblock {\em Proceedings of the ACM on Human-Computer Interaction}, 5(CSCW2):1--29, 2021.

\bibitem{bandy2021problematic}
Jack Bandy.
\newblock Problematic machine behavior: A systematic literature review of algorithm audits.
\newblock {\em Proceedings of the acm on human-computer interaction}, 5(CSCW1):1--34, 2021.

\bibitem{scott2011auditing}
Peter~R Scott and J~Mike Jacka.
\newblock {\em Auditing social media: A governance and risk guide}.
\newblock John Wiley \& Sons, 2011.

\bibitem{goel2012does}
Sharad Goel, Jake Hofman, and M~Sirer.
\newblock Who does what on the web: A large-scale study of browsing behavior.
\newblock In {\em Proceedings of the International AAAI Conference on web and Social Media}, volume~6, pages 130--137, 2012.

\bibitem{kumar2010characterization}
Ravi Kumar and Andrew Tomkins.
\newblock A characterization of online browsing behavior.
\newblock In {\em Proceedings of the 19th international conference on World wide web}, pages 561--570, 2010.

\bibitem{bucklin2003model}
Randolph~E Bucklin and Catarina Sismeiro.
\newblock A model of web site browsing behavior estimated on clickstream data.
\newblock {\em Journal of marketing research}, 40(3):249--267, 2003.

\bibitem{metzger2018benefits}
Miriam~J Metzger, Christo Wilson, and Ben~Y Zhao.
\newblock Benefits of browsing? the prevalence, nature, and effects of profile consumption behavior in social network sites.
\newblock {\em Journal of Computer-Mediated Communication}, 23(2):72--89, 2018.

\bibitem{rand2015agent}
William Rand, Jeffrey Herrmann, Brandon Schein, and Ne{\v{z}}a Vodopivec.
\newblock An agent-based model of urgent diffusion in social media.
\newblock {\em Journal of Artificial Societies and Social Simulation}, 18(2):1, 2015.

\bibitem{toriumi2012people}
Fujio Toriumi, Hitoshi Yamamoto, and Isamu Okada.
\newblock Why do people use social media? agent-based simulation and population dynamics analysis of the evolution of cooperation in social media.
\newblock In {\em 2012 IEEE/WIC/ACM International Conferences on Web Intelligence and Intelligent Agent Technology}, volume~2, pages 43--50. IEEE, 2012.

\bibitem{plikynas2015agent}
Darius Plikynas, Aistis Raudys, and Sarunas Raudys.
\newblock Agent-based modelling of excitation propagation in social media groups.
\newblock {\em Journal of Experimental \& Theoretical Artificial Intelligence}, 27(4):373--388, 2015.

\bibitem{kaligotla2015agent}
Chaitanya Kaligotla, Enver Y{\"u}cesan, and Stephen~E Chick.
\newblock An agent based model of spread of competing rumors through online interactions on social media.
\newblock In {\em 2015 winter simulation conference (WSC)}, pages 3985--3996. IEEE, 2015.

\bibitem{zollo2024understanding}
Fabiana Zollo, Andrea Baronchelli, Cornelia Betsch, Marco Delmastro, and Walter Quattrociocchi.
\newblock Understanding the complex links between social media and health behaviour.
\newblock {\em bmj}, 385, 2024.

\bibitem{wagner2023independence}
Michael~W Wagner.
\newblock Independence by permission.
\newblock {\em Science}, 381(6656):388--391, 2023.

\bibitem{verma2014editorial}
Inder~M Verma.
\newblock Editorial expression of concern: Experimental evidence of massive-scale emotional contagion through social networks.
\newblock {\em Proceedings of the National Academy of Sciences of the United States of America}, 111(29):10779--10779, 2014.

\bibitem{esteve2017business}
Asunci{\'o}n Esteve.
\newblock The business of personal data: Google, facebook, and privacy issues in the eu and the usa.
\newblock {\em International Data Privacy Law}, 7(1):36--47, 2017.

\bibitem{fagiolo2006empirical}
Giorgio Fagiolo, Paul Windrum, and Alessio Moneta.
\newblock Empirical validation of agent-based models: A critical survey.
\newblock Technical report, LEM Working Paper Series, 2006.

\bibitem{manzo2014potentialities}
Gianluca Manzo and Toby Matthews.
\newblock Potentialities and limitations of agent-based simulations.
\newblock {\em Revue fran{\c{c}}aise de sociologie}, 55(4):653--688, 2014.

\bibitem{windrum2007empirical}
Paul Windrum, Giorgio Fagiolo, and Alessio Moneta.
\newblock Empirical validation of agent-based models: Alternatives and prospects.
\newblock {\em Journal of Artificial Societies and Social Simulation}, 10(2):8, 2007.

\bibitem{ghani2019social}
Norjihan~Abdul Ghani, Suraya Hamid, Ibrahim Abaker~Targio Hashem, and Ejaz Ahmed.
\newblock Social media big data analytics: A survey.
\newblock {\em Computers in Human behavior}, 101:417--428, 2019.

\bibitem{kitsak2010identification}
Maksim Kitsak, Lazaros~K Gallos, Shlomo Havlin, Fredrik Liljeros, Lev Muchnik, H~Eugene Stanley, and Hern{\'a}n~A Makse.
\newblock Identification of influential spreaders in complex networks.
\newblock {\em Nature physics}, 6(11):888--893, 2010.

\bibitem{corbyn2012facebook}
Zoe Corbyn.
\newblock Facebook experiment boosts us voter turnout.
\newblock {\em Nature}, 12, 2012.

\bibitem{menczer2020attention}
Filippo Menczer and Thomas Hills.
\newblock The attention economy.
\newblock {\em Scientific American}, 323(6):54--61, 2020.

\bibitem{ciampaglia2015production}
Giovanni~Luca Ciampaglia, Alessandro Flammini, and Filippo Menczer.
\newblock The production of information in the attention economy.
\newblock {\em Scientific reports}, 5(1):9452, 2015.

\bibitem{goldhaber1997attention}
Michael~H Goldhaber.
\newblock The attention economy and the net.
\newblock {\em First Monday}, 1997.

\bibitem{pei2014searching}
Sen Pei, Lev Muchnik, Jos{\'e}~S Andrade, Jr, Zhiming Zheng, and Hern{\'a}n~A Makse.
\newblock Searching for superspreaders of information in real-world social media.
\newblock {\em Scientific reports}, 4(1):5547, 2014.

\bibitem{zareie2021minimizing}
Ahmad Zareie and Rizos Sakellariou.
\newblock Minimizing the spread of misinformation in online social networks: A survey.
\newblock {\em Journal of Network and Computer Applications}, 186:103094, 2021.

\bibitem{an2013fragmented}
Jisun An, Daniele Quercia, and Jon Crowcroft.
\newblock Fragmented social media: a look into selective exposure to political news.
\newblock In {\em Proceedings of the 22nd international conference on world wide web}, pages 51--52, 2013.

\bibitem{Davenport2001}
Thomas~H. Davenport and John~C. Beck.
\newblock The attention economy.
\newblock {\em Ubiquity}, 2001(May):1, May 2001.

\bibitem{Falkinger2008}
Josef Falkinger.
\newblock Limited attention as a scarce resource in information‐rich economies.
\newblock {\em The Economic Journal}, 118(532):1596–1620, September 2008.

\bibitem{spohr2017fake}
Dominic Spohr.
\newblock Fake news and ideological polarization: Filter bubbles and selective exposure on social media.
\newblock {\em Business information review}, 34(3):150--160, 2017.

\bibitem{anderson2012competition}
Simon~P Anderson and Andr{\'e} De~Palma.
\newblock Competition for attention in the information (overload) age.
\newblock {\em The RAND Journal of Economics}, 43(1):1--25, 2012.

\bibitem{Hargittai2012}
Eszter Hargittai, W.~Russell Neuman, and Olivia Curry.
\newblock Taming the information tide: Perceptions of information overload in the american home.
\newblock {\em The Information Society}, 28(3):161–173, May 2012.

\bibitem{Bawden2020}
David Bawden and Lyn Robinson.
\newblock Information overload: An introduction, June 2020.

\bibitem{cinelli2020selective}
Matteo Cinelli, Emanuele Brugnoli, Ana~Lucia Schmidt, Fabiana Zollo, Walter Quattrociocchi, and Antonio Scala.
\newblock Selective exposure shapes the facebook news diet.
\newblock {\em PloS one}, 15(3):e0229129, 2020.

\bibitem{Waller2012}
Aaron~David Waller and Gillian Ragsdell.
\newblock The impact of e‐mail on work‐life balance.
\newblock {\em Aslib Proceedings}, 64(2):154–177, March 2012.

\bibitem{Tufekci2013}
Zeynep Tufekci.
\newblock “not this one”: Social movements, the attention economy, and microcelebrity networked activism.
\newblock {\em American Behavioral Scientist}, 57(7):848–870, March 2013.

\bibitem{Bhargava2020}
Vikram~R. Bhargava and Manuel Velasquez.
\newblock Ethics of the attention economy: The problem of social media addiction.
\newblock {\em Business Ethics Quarterly}, 31(3):321–359, October 2020.

\bibitem{bessi2016users}
Alessandro Bessi, Fabiana Zollo, Michela Del~Vicario, Michelangelo Puliga, Antonio Scala, Guido Caldarelli, Brian Uzzi, and Walter Quattrociocchi.
\newblock Users polarization on facebook and youtube.
\newblock {\em PloS one}, 11(8):e0159641, 2016.

\bibitem{cheng2008statistics}
Xu~Cheng, Cameron Dale, and Jiangchuan Liu.
\newblock Statistics and social network of youtube videos.
\newblock In {\em 2008 16th Interntional Workshop on Quality of Service}, pages 229--238. IEEE, 2008.

\bibitem{Sangiorgio2024}
Emanuele Sangiorgio, Matteo Cinelli, Roy Cerqueti, and Walter Quattrociocchi.
\newblock Followers do not dictate the virality of news outlets on social media.
\newblock {\em PNAS Nexus}, 3(7), June 2024.

\bibitem{lerman2007social}
Kristina Lerman.
\newblock Social information processing in news aggregation.
\newblock {\em IEEE internet computing}, 11(6):16--28, 2007.

\bibitem{Etta2023}
Gabriele Etta, Emanuele Sangiorgio, Niccolò Di~Marco, Michele Avalle, Antonio Scala, Matteo Cinelli, and Walter Quattrociocchi.
\newblock Characterizing engagement dynamics across topics on facebook.
\newblock {\em PLOS ONE}, 18(6):e0286150, June 2023.

\bibitem{Robertson2023}
Claire~E. Robertson, Nicolas Pr\"{o}llochs, Kaoru Schwarzenegger, Philip P\"{a}rnamets, Jay~J. Van~Bavel, and Stefan Feuerriegel.
\newblock Negativity drives online news consumption.
\newblock {\em Nature Human Behaviour}, 7(5):812–822, March 2023.

\bibitem{torricelli2023does}
Maddalena Torricelli, Max Falkenberg, Alessandro Galeazzi, Fabiana Zollo, Walter Quattrociocchi, and Andrea Baronchelli.
\newblock How does extreme weather impact the climate change discourse? insights from the twitter discussion on hurricanes.
\newblock {\em Plos Climate}, 2(11):e0000277, 2023.

\bibitem{gallotti2020assessing}
Riccardo Gallotti, Francesco Valle, Nicola Castaldo, Pierluigi Sacco, and Manlio De~Domenico.
\newblock Assessing the risks of ‘infodemics’ in response to covid-19 epidemics.
\newblock {\em Nature human behaviour}, 4(12):1285--1293, 2020.

\bibitem{bessi2015viral}
Alessandro Bessi, Fabio Petroni, Michela Del~Vicario, Fabiana Zollo, Aris Anagnostopoulos, Antonio Scala, Guido Caldarelli, and Walter Quattrociocchi.
\newblock Viral misinformation: The role of homophily and polarization.
\newblock In {\em Proceedings of the 24th international conference on World Wide Web}, pages 355--356, 2015.

\bibitem{bak2022combining}
Joseph~B Bak-Coleman, Ian Kennedy, Morgan Wack, Andrew Beers, Joseph~S Schafer, Emma~S Spiro, Kate Starbird, and Jevin~D West.
\newblock Combining interventions to reduce the spread of viral misinformation.
\newblock {\em Nature Human Behaviour}, 6(10):1372--1380, 2022.

\bibitem{kim2018they}
Ji~Won Kim.
\newblock They liked and shared: Effects of social media virality metrics on perceptions of message influence and behavioral intentions.
\newblock {\em Computers in Human Behavior}, 84:153--161, 2018.

\bibitem{mousavi2022effective}
Maryam Mousavi, Hasan Davulcu, Mohsen Ahmadi, Robert Axelrod, Richard Davis, and Scott Atran.
\newblock Effective messaging on social media: What makes online content go viral?
\newblock In {\em Proceedings of the ACM Web Conference 2022}, pages 2957--2966, 2022.

\bibitem{nikolinakou2018viral}
Angeliki Nikolinakou and Karen~Whitehill King.
\newblock Viral video ads: Emotional triggers and social media virality.
\newblock {\em Psychology \& marketing}, 35(10):715--726, 2018.

\bibitem{fernandez2024analysing}
Miriam Fernandez, Alejandro Bellog{\'\i}n, and Iv{\'a}n Cantador.
\newblock Analysing the effect of recommendation algorithms on the spread of misinformation.
\newblock In {\em Proceedings of the 16th ACM Web Science Conference}, pages 159--169, 2024.

\bibitem{guerini2011exploring}
Marco Guerini, Carlo Strapparava, and Gozde Ozbal.
\newblock Exploring text virality in social networks.
\newblock In {\em proceedings of the international AAAI conference on web and social media}, volume~5, pages 506--509, 2011.

\bibitem{han2020importance}
Yue Han, Theodoros Lappas, and Gaurav Sabnis.
\newblock The importance of interactions between content characteristics and creator characteristics for studying virality in social media.
\newblock {\em Information Systems Research}, 31(2):576--588, 2020.

\bibitem{di2024users}
Niccol{\`o} Di~Marco, Matteo Cinelli, Shayan Alipour, and Walter Quattrociocchi.
\newblock Users volatility on reddit and voat.
\newblock {\em IEEE Transactions on Computational Social Systems}, 2024.

\bibitem{sangiorgio2025evaluating}
Emanuele Sangiorgio, Niccol{\`o} Di~Marco, Gabriele Etta, Matteo Cinelli, Roy Cerqueti, and Walter Quattrociocchi.
\newblock Evaluating the effect of viral posts on social media engagement.
\newblock {\em Scientific Reports}, 15(1):639, 2025.

\bibitem{stieglitz2013emotions}
Stefan Stieglitz and Linh Dang-Xuan.
\newblock Emotions and information diffusion in social media—sentiment of microblogs and sharing behavior.
\newblock {\em Journal of management information systems}, 29(4):217--248, 2013.

\bibitem{Amini2025}
Arash Amini, Yigit~Ege Bayiz, Eun-Ju Lee, Zeynep Somer-Topcu, Radu Marculescu, and Ufuk Topcu.
\newblock How media competition fuels the spread of misinformation.
\newblock {\em Science Advances}, 11(25), June 2025.

\bibitem{ecker2022psychological}
Ullrich~KH Ecker, Stephan Lewandowsky, John Cook, Philipp Schmid, Lisa~K Fazio, Nadia Brashier, Panayiota Kendeou, Emily~K Vraga, and Michelle~A Amazeen.
\newblock The psychological drivers of misinformation belief and its resistance to correction.
\newblock {\em Nature Reviews Psychology}, 1(1):13--29, 2022.

\bibitem{briand2021infodemics}
Sylvie~C Briand, Matteo Cinelli, Tim Nguyen, Rosamund Lewis, Dimitri Prybylski, Carlo~M Valensise, Vittoria Colizza, Alberto~Eugenio Tozzi, Nicola Perra, Andrea Baronchelli, et~al.
\newblock Infodemics: A new challenge for public health.
\newblock {\em Cell}, 184(25):6010--6014, 2021.

\bibitem{mccombs1972agenda}
Maxwell~E McCombs and Donald~L Shaw.
\newblock The agenda-setting function of mass media.
\newblock {\em Public opinion quarterly}, 36(2):176--187, 1972.

\bibitem{gilardi2022social}
Fabrizio Gilardi, Theresa Gessler, Ma{\"e}l Kubli, and Stefan M{\"u}ller.
\newblock Social media and political agenda setting.
\newblock {\em Political communication}, 39(1):39--60, 2022.

\bibitem{yang2016social}
Xinxin Yang, Bo-Chiuan Chen, Mrinmoy Maity, and Emilio Ferrara.
\newblock Social politics: Agenda setting and political communication on social media.
\newblock In {\em Social Informatics: 8th International Conference, SocInfo 2016, Bellevue, WA, USA, November 11-14, 2016, Proceedings, Part I 8}, pages 330--344. Springer, 2016.

\bibitem{quattrociocchi2014opinion}
Walter Quattrociocchi, Guido Caldarelli, and Antonio Scala.
\newblock Opinion dynamics on interacting networks: media competition and social influence.
\newblock {\em Scientific reports}, 4(1):4938, 2014.

\bibitem{kim2025agenda}
Rachel~M Kim and Ashton Anderson.
\newblock The agenda-setting function of social media.
\newblock In {\em Proceedings of the ACM on Web Conference 2025}, pages 601--613, 2025.

\bibitem{meraz2009there}
Sharon Meraz.
\newblock Is there an elite hold? traditional media to social media agenda setting influence in blog networks.
\newblock {\em Journal of computer-mediated communication}, 14(3):682--707, 2009.

\bibitem{ding2019social}
Keyan Ding, Ronggang Wang, and Shiqi Wang.
\newblock Social media popularity prediction: A multiple feature fusion approach with deep neural networks.
\newblock In {\em Proceedings of the 27th ACM International Conference on Multimedia}, pages 2682--2686, 2019.

\bibitem{kong2018exploring}
Qingchao Kong, Wenji Mao, Guandan Chen, and Daniel Zeng.
\newblock Exploring trends and patterns of popularity stage evolution in social media.
\newblock {\em IEEE Transactions on Systems, Man, and Cybernetics: Systems}, 50(10):3817--3827, 2018.

\bibitem{trunfio2021conceptualising}
Mariapina Trunfio and Simona Rossi.
\newblock Conceptualising and measuring social media engagement: A systematic literature review.
\newblock {\em Italian Journal of Marketing}, 2021(3):267--292, 2021.

\bibitem{zaccaria2019poprank}
Andrea Zaccaria, Michela Del~Vicario, Walter Quattrociocchi, Antonio Scala, and Luciano Pietronero.
\newblock Poprank: Ranking pages’ impact and users’ engagement on facebook.
\newblock {\em PloS one}, 14(1):e0211038, 2019.

\bibitem{moehring2023providing}
Alex Moehring, Avinash Collis, Kiran Garimella, M~Amin Rahimian, Sinan Aral, and Dean Eckles.
\newblock Providing normative information increases intentions to accept a covid-19 vaccine.
\newblock {\em Nature Communications}, 14(1):126, 2023.

\bibitem{bakshy2015exposure}
Eytan Bakshy, Solomon Messing, and Lada~A Adamic.
\newblock Exposure to ideologically diverse news and opinion on facebook.
\newblock {\em Science}, 348(6239):1130--1132, 2015.

\bibitem{weng2018attention}
Lilian Weng, M{\'a}rton Karsai, Nicola Perra, Filippo Menczer, and Alessandro Flammini.
\newblock Attention on weak ties in social and communication networks.
\newblock {\em Complex spreading phenomena in social systems: Influence and contagion in real-world social networks}, pages 213--228, 2018.

\bibitem{gonccalves2011modeling}
Bruno Gon{\c{c}}alves, Nicola Perra, and Alessandro Vespignani.
\newblock Modeling users' activity on twitter networks: Validation of dunbar's number.
\newblock {\em PloS one}, 6(8):e22656, 2011.

\bibitem{Loru2024agenda}
Edoardo Loru, Alessandro Galeazzi, Anita Bonetti, Emanuele Sangiorgio, Niccolò Di~Marco, Matteo Cinelli, Andrea Baronchelli, and Walter Quattrociocchi.
\newblock Who sets the agenda on social media? ideology and polarization in online debates, 2024.

\bibitem{falkenberg2022growing}
Max Falkenberg, Alessandro Galeazzi, Maddalena Torricelli, Niccol{\`o} Di~Marco, Francesca Larosa, Madalina Sas, Amin Mekacher, Warren Pearce, Fabiana Zollo, Walter Quattrociocchi, et~al.
\newblock Growing polarization around climate change on social media.
\newblock {\em Nature Climate Change}, 12(12):1114--1121, 2022.

\bibitem{covington2016deep}
Paul Covington, Jay Adams, and Emre Sargin.
\newblock Deep neural networks for youtube recommendations.
\newblock In {\em Proceedings of the 10th ACM conference on recommender systems}, pages 191--198, 2016.

\bibitem{biega2018equity}
Asia~J Biega, Krishna~P Gummadi, and Gerhard Weikum.
\newblock Equity of attention: Amortizing individual fairness in rankings.
\newblock In {\em The 41st international acm sigir conference on research \& development in information retrieval}, pages 405--414, 2018.

\bibitem{bakshy2014designing}
Eytan Bakshy, Dean Eckles, and Michael~S Bernstein.
\newblock Designing and deploying online field experiments.
\newblock In {\em Proceedings of the 23rd international conference on World wide web}, pages 283--292, 2014.

\bibitem{quattrociocchi2017inside}
Walter Quattrociocchi.
\newblock Inside the echo chamber.
\newblock {\em Scientific American}, 316(4):60--63, 2017.

\bibitem{mahmoudi2024echo}
Amin Mahmoudi, Dariusz Jemielniak, and Leon Ciechanowski.
\newblock Echo chambers in online social networks: A systematic literature review.
\newblock {\em IEEE Access}, 12:9594--9620, 2024.

\bibitem{dubois2018echo}
Elizabeth Dubois and Grant Blank.
\newblock The echo chamber is overstated: the moderating effect of political interest and diverse media.
\newblock {\em Information, communication \& society}, 21(5):729--745, 2018.

\bibitem{falkenberg2024patterns}
Max Falkenberg, Fabiana Zollo, Walter Quattrociocchi, J{\"u}rgen Pfeffer, and Andrea Baronchelli.
\newblock Patterns of partisan toxicity and engagement reveal the common structure of online political communication across countries.
\newblock {\em Nature Communications}, 15(1):9560, 2024.

\bibitem{gaines2009typing}
Brian~J Gaines and Jeffery~J Mondak.
\newblock Typing together? clustering of ideological types in online social networks.
\newblock {\em Journal of Information Technology \& Politics}, 6(3-4):216--231, 2009.

\bibitem{munger2019supply}
Kevin Munger and Joseph Phillips.
\newblock A supply and demand framework for youtube politics.
\newblock {\em preprint}, 2019.

\bibitem{hartmann2025systematic}
David Hartmann, Sonja~Mei Wang, Lena Pohlmann, and Bettina Berendt.
\newblock A systematic review of echo chamber research: Comparative analysis of conceptualizations, operationalizations, and varying outcomes.
\newblock {\em Journal of Computational Social Science}, 8(2):52, 2025.

\bibitem{amendola2024towards}
Miriam Amendola, Danilo Cavaliere, Carmen De~Maio, Giuseppe Fenza, and Vincenzo Loia.
\newblock Towards echo chamber assessment by employing aspect-based sentiment analysis and gdm consensus metrics.
\newblock {\em Online Social Networks and Media}, 39:100276, 2024.

\bibitem{williams2015network}
Hywel~TP Williams, James~R McMurray, Tim Kurz, and F~Hugo Lambert.
\newblock Network analysis reveals open forums and echo chambers in social media discussions of climate change.
\newblock {\em Global environmental change}, 32:126--138, 2015.

\bibitem{Zollo2015}
Fabiana Zollo, Petra~Kralj Novak, Michela Del~Vicario, Alessandro Bessi, Igor Mozetič, Antonio Scala, Guido Caldarelli, and Walter Quattrociocchi.
\newblock Emotional dynamics in the age of misinformation.
\newblock {\em PLOS ONE}, 10(9):e0138740, September 2015.

\bibitem{brugnoli2019recursive}
Emanuele Brugnoli, Matteo Cinelli, Walter Quattrociocchi, and Antonio Scala.
\newblock Recursive patterns in online echo chambers.
\newblock {\em Scientific Reports}, 9(1):20118, 2019.

\bibitem{sasahara2021social}
Kazutoshi Sasahara, Wen Chen, Hao Peng, Giovanni~Luca Ciampaglia, Alessandro Flammini, and Filippo Menczer.
\newblock Social influence and unfollowing accelerate the emergence of echo chambers.
\newblock {\em Journal of Computational Social Science}, 4(1):381--402, 2021.

\bibitem{Rathje2021}
Steve Rathje, Jay~J. Van~Bavel, and Sander van~der Linden.
\newblock Out-group animosity drives engagement on social media.
\newblock {\em Proceedings of the National Academy of Sciences}, 118(26), June 2021.

\bibitem{choi2020rumor}
Daejin Choi, Selin Chun, Hyunchul Oh, Jinyoung Han, and Ted~“Taekyoung” Kwon.
\newblock Rumor propagation is amplified by echo chambers in social media.
\newblock {\em Scientific reports}, 10(1):310, 2020.

\bibitem{robertson2023negativity}
Claire~E Robertson, Nicolas Pr{\"o}llochs, Kaoru Schwarzenegger, Philip P{\"a}rnamets, Jay~J Van~Bavel, and Stefan Feuerriegel.
\newblock Negativity drives online news consumption.
\newblock {\em Nature human behaviour}, 7(5):812--822, 2023.

\bibitem{guess2020exposure}
Andrew~M Guess, Brendan Nyhan, and Jason Reifler.
\newblock Exposure to untrustworthy websites in the 2016 us election.
\newblock {\em Nature human behaviour}, 4(5):472--480, 2020.

\bibitem{pennycook2019lazy}
Gordon Pennycook and David~G Rand.
\newblock Lazy, not biased: Susceptibility to partisan fake news is better explained by lack of reasoning than by motivated reasoning.
\newblock {\em Cognition}, 188:39--50, 2019.

\bibitem{DelVicario2016}
Michela Del~Vicario, Alessandro Bessi, Fabiana Zollo, Fabio Petroni, Antonio Scala, Guido Caldarelli, H.~Eugene Stanley, and Walter Quattrociocchi.
\newblock The spreading of misinformation online.
\newblock {\em Proceedings of the National Academy of Sciences}, 113(3):554–559, January 2016.

\bibitem{Zollo2017}
Fabiana Zollo, Alessandro Bessi, Michela Del~Vicario, Antonio Scala, Guido Caldarelli, Louis Shekhtman, Shlomo Havlin, and Walter Quattrociocchi.
\newblock Debunking in a world of tribes.
\newblock {\em PLOS ONE}, 12(7):e0181821, July 2017.

\bibitem{ciampaglia2018research}
Giovanni~Luca Ciampaglia, Alexios Mantzarlis, Gregory Maus, and Filippo Menczer.
\newblock Research challenges of digital misinformation: Toward a trustworthy web.
\newblock {\em AI Magazine}, 39(1):65--74, 2018.

\bibitem{lazer2018science}
David~MJ Lazer, Matthew~A Baum, Yochai Benkler, Adam~J Berinsky, Kelly~M Greenhill, Filippo Menczer, Miriam~J Metzger, Brendan Nyhan, Gordon Pennycook, David Rothschild, et~al.
\newblock The science of fake news.
\newblock {\em Science}, 359(6380):1094--1096, 2018.

\bibitem{juul2021comparing}
Jonas~L Juul and Johan Ugander.
\newblock Comparing information diffusion mechanisms by matching on cascade size.
\newblock {\em Proceedings of the National Academy of Sciences}, 118(46):e2100786118, 2021.

\bibitem{martel2021you}
Cameron Martel, Mohsen Mosleh, and David~G Rand.
\newblock You’re definitely wrong, maybe: Correction style has minimal effect on corrections of misinformation online.
\newblock {\em Media and Communication}, 9(1):120--133, 2021.

\bibitem{martel2023misinformation}
Cameron Martel and David~G Rand.
\newblock Misinformation warning labels are widely effective: A review of warning effects and their moderating features.
\newblock {\em Current Opinion in Psychology}, 54:101710, 2023.

\bibitem{jia2022understanding}
Chenyan Jia, Alexander Boltz, Angie Zhang, Anqing Chen, and Min~Kyung Lee.
\newblock Understanding effects of algorithmic vs. community label on perceived accuracy of hyper-partisan misinformation.
\newblock {\em Proceedings of the ACM on Human-Computer Interaction}, 6(CSCW2):1--27, 2022.

\bibitem{maertens2021long}
Rakoen Maertens, Jon Roozenbeek, Melisa Basol, and Sander van~der Linden.
\newblock Long-term effectiveness of inoculation against misinformation: Three longitudinal experiments.
\newblock {\em Journal of Experimental Psychology: Applied}, 27(1):1, 2021.

\bibitem{lewandowsky2012misinformation}
Stephan Lewandowsky, Ullrich~KH Ecker, Colleen~M Seifert, Norbert Schwarz, and John Cook.
\newblock Misinformation and its correction: Continued influence and successful debiasing.
\newblock {\em Psychological science in the public interest}, 13(3):106--131, 2012.

\bibitem{iizuka2022impact}
Ryusuke Iizuka, Fujio Toriumi, Mao Nishiguchi, Masanori Takano, and Mitsuo Yoshida.
\newblock Impact of correcting misinformation on social disruption.
\newblock {\em Plos one}, 17(4):e0265734, 2022.

\bibitem{chan2023meta}
Man-pui~Sally Chan and Dolores Albarracín.
\newblock A meta-analysis of correction effects in science-relevant misinformation.
\newblock {\em Nature Human Behaviour}, 7(9):1514--1525, 2023.

\bibitem{taylor2023identity}
Sean~J Taylor, Lev Muchnik, Madhav Kumar, and Sinan Aral.
\newblock Identity effects in social media.
\newblock {\em Nature Human Behaviour}, 7(1):27--37, 2023.

\bibitem{Walter2019}
Nathan Walter, Jonathan Cohen, R.~Lance Holbert, and Yasmin Morag.
\newblock Fact-checking: A meta-analysis of what works and for whom.
\newblock {\em Political Communication}, 37(3):350–375, October 2019.

\bibitem{Porter2021}
Ethan Porter and Thomas~J. Wood.
\newblock The global effectiveness of fact-checking: Evidence from simultaneous experiments in argentina, nigeria, south africa, and the united kingdom.
\newblock {\em Proceedings of the National Academy of Sciences}, 118(37), September 2021.

\bibitem{SwireThompson2020}
Briony Swire-Thompson, Joseph DeGutis, and David Lazer.
\newblock Searching for the backfire effect: Measurement and design considerations.
\newblock {\em Journal of Applied Research in Memory and Cognition}, 9(3):286–299, September 2020.

\bibitem{swire2020searching}
Briony Swire-Thompson, Joseph DeGutis, and David Lazer.
\newblock Searching for the backfire effect: Measurement and design considerations.
\newblock {\em Journal of applied research in memory and cognition}, 9(3):286--299, 2020.

\bibitem{nyhan2021backfire}
Brendan Nyhan.
\newblock Why the backfire effect does not explain the durability of political misperceptions.
\newblock {\em Proceedings of the National Academy of Sciences}, 118(15):e1912440117, 2021.

\bibitem{van2022misinformation}
Sander Van Der~Linden.
\newblock Misinformation: susceptibility, spread, and interventions to immunize the public.
\newblock {\em Nature medicine}, 28(3):460--467, 2022.

\bibitem{basol2021towards}
Melisa Basol, Jon Roozenbeek, Manon Berriche, Fatih Uenal, William~P McClanahan, and Sander van~der Linden.
\newblock Towards psychological herd immunity: Cross-cultural evidence for two prebunking interventions against covid-19 misinformation.
\newblock {\em Big Data \& Society}, 8(1):20539517211013868, 2021.

\bibitem{van2024inoculation}
Sander van~der Linden and Jon Roozenbeek.
\newblock “inoculation” to resist misinformation.
\newblock {\em Jama}, 331(22):1961--1962, 2024.

\bibitem{traberg2022psychological}
Cecilie~S Traberg, Jon Roozenbeek, and Sander van~der Linden.
\newblock Psychological inoculation against misinformation: Current evidence and future directions.
\newblock {\em The ANNALS of the American Academy of Political and Social Science}, 700(1):136--151, 2022.

\bibitem{van2020psychological}
Sander Van~der Linden, Jon Roozenbeek, et~al.
\newblock Psychological inoculation against fake news.
\newblock {\em The psychology of fake news: Accepting, sharing, and correcting misinformation}, pages 147--169, 2020.

\bibitem{pacheco2021uncovering}
Diogo Pacheco, Pik-Mai Hui, Christopher Torres-Lugo, Bao~Tran Truong, Alessandro Flammini, and Filippo Menczer.
\newblock Uncovering coordinated networks on social media: methods and case studies.
\newblock In {\em Proceedings of the international AAAI conference on web and social media}, volume~15, pages 455--466, 2021.

\bibitem{Tardelli2024temporal}
Serena Tardelli, Leonardo Nizzoli, Maurizio Tesconi, Mauro Conti, Preslav Nakov, Giovanni Da~San~Martino, and Stefano Cresci.
\newblock Temporal dynamics of coordinated online behavior: Stability, archetypes, and influence.
\newblock {\em Proceedings of the National Academy of Sciences}, 121(20), May 2024.

\bibitem{Cinelli2022}
Matteo Cinelli, Stefano Cresci, Walter Quattrociocchi, Maurizio Tesconi, and Paola Zola.
\newblock Coordinated inauthentic behavior and information spreading on twitter.
\newblock {\em Decision Support Systems}, 160:113819, September 2022.

\bibitem{hristakieva2022spread}
Kristina Hristakieva, Stefano Cresci, Giovanni Da~San~Martino, Mauro Conti, and Preslav Nakov.
\newblock The spread of propaganda by coordinated communities on social media.
\newblock In {\em Proceedings of the 14th ACM web science conference 2022}, pages 191--201, 2022.

\bibitem{weber2021amplifying}
Derek Weber and Frank Neumann.
\newblock Amplifying influence through coordinated behaviour in social networks.
\newblock {\em Social Network Analysis and Mining}, 11(1), 2021.

\bibitem{ferrara2016rise}
Emilio Ferrara, Onur Varol, Clayton Davis, Filippo Menczer, and Alessandro Flammini.
\newblock The rise of social bots.
\newblock {\em Communications of the ACM}, 59(7):96--104, 2016.

\bibitem{Cresci2020decade}
Stefano Cresci.
\newblock A decade of social bot detection.
\newblock {\em Communications of the ACM}, 63(10):72–83, September 2020.

\bibitem{Cheng2017anyone}
Justin Cheng, Michael Bernstein, Cristian Danescu-Niculescu-Mizil, and Jure Leskovec.
\newblock Anyone can become a troll: Causes of trolling behavior in online discussions.
\newblock In {\em Proceedings of the 2017 ACM Conference on Computer Supported Cooperative Work and Social Computing}, CSCW ’17, page 1217–1230. ACM, February 2017.

\bibitem{Nizzoli2021}
Leonardo Nizzoli, Serena Tardelli, Marco Avvenuti, Stefano Cresci, and Maurizio Tesconi.
\newblock Coordinated behavior on social media in 2019 uk general election.
\newblock {\em Proceedings of the International AAAI Conference on Web and Social Media}, 15:443–454, May 2021.

\bibitem{badawy2018analyzing}
Adam Badawy, Emilio Ferrara, and Kristina Lerman.
\newblock Analyzing the digital traces of political manipulation: The 2016 russian interference twitter campaign.
\newblock In {\em 2018 IEEE/ACM international conference on advances in social networks analysis and mining (ASONAM)}, pages 258--265. IEEE, 2018.

\bibitem{Shu2020}
Kai Shu, Suhang Wang, Dongwon Lee, and Huan Liu.
\newblock {\em Mining Disinformation and Fake News: Concepts, Methods, and Recent Advancements}, page 1–19.
\newblock Springer International Publishing, 2020.

\bibitem{Starbird2019disinformation}
Kate Starbird, Ahmer Arif, and Tom Wilson.
\newblock Disinformation as collaborative work: Surfacing the participatory nature of strategic information operations.
\newblock {\em Proceedings of the ACM on Human-Computer Interaction}, 3(CSCW):1–26, November 2019.

\bibitem{stella2018bots}
Massimo Stella, Emilio Ferrara, and Manlio De~Domenico.
\newblock Bots increase exposure to negative and inflammatory content in online social systems.
\newblock {\em Proceedings of the National Academy of Sciences}, 115(49):12435--12440, 2018.

\bibitem{luceri2019red}
Luca Luceri, Ashok Deb, Adam Badawy, and Emilio Ferrara.
\newblock Red bots do it better: Comparative analysis of social bot partisan behavior.
\newblock In {\em Companion proceedings of the 2019 world wide web conference}, pages 1007--1012, 2019.

\bibitem{keller2020political}
Franziska~B. Keller, David Schoch, Sebastian Stier, and JungHwan Yang.
\newblock Political {Astroturfing} on {Twitter}: How to {Coordinate} a {Disinformation} {Campaign}.
\newblock {\em Political Communication}, 37(2):256--280, 2020.

\bibitem{schoch2022coordination}
David Schoch, Franziska~B. Keller, Sebastian Stier, and JungHwan Yang.
\newblock Coordination patterns reveal online political astroturfing across the world.
\newblock {\em Scientific Reports}, 12(1), 2022.

\bibitem{Ratkiewicz2021detecting}
Jacob Ratkiewicz, Michael Conover, Mark Meiss, Bruno Goncalves, Alessandro Flammini, and Filippo Menczer.
\newblock Detecting and tracking political abuse in social media.
\newblock {\em Proceedings of the International AAAI Conference on Web and Social Media}, 5(1):297–304, August 2021.

\bibitem{ratkiewicz2011truthy}
Jacob Ratkiewicz, Michael Conover, Mark Meiss, Bruno Gon{\c{c}}alves, Snehal Patil, Alessandro Flammini, and Filippo Menczer.
\newblock Truthy: mapping the spread of astroturf in microblog streams.
\newblock In {\em Proceedings of the 20th international conference companion on World wide web}, pages 249--252, 2011.

\bibitem{Cinelli2019information}
M~Cinelli, M~Conti, L~Finos, F~Grisolia, P~Kralj Novak, A~Peruzzi, M~Tesconi, F~Zollo, and W~Quattrociocchi.
\newblock (mis)information operations: An integrated perspective.
\newblock {\em Journal of Information Warfare}, 18(3):83--98, 2019.

\bibitem{bennett2012logic}
W~Lance Bennett and Alexandra Segerberg.
\newblock The logic of connective action: Digital media and the personalization of contentious politics.
\newblock {\em Information, communication \& society}, 15(5):739--768, 2012.

\bibitem{weller2013twitter}
Katrin Weller, Axel Bruns, Jean~Elizabeth Burgess, Merja Mahrt, and Cornelius Puschmann.
\newblock {\em Twitter and society}.
\newblock Peter Lang New York, 2013.

\bibitem{castillo2014characterizing}
Carlos Castillo, Mohammed El-Haddad, J{\"u}rgen Pfeffer, and Matt Stempeck.
\newblock Characterizing the life cycle of online news stories using social media reactions.
\newblock In {\em Proceedings of the 17th ACM conference on Computer supported cooperative work \& social computing}, pages 211--223, 2014.

\bibitem{orabi2020detection}
Mariam Orabi, Djedjiga Mouheb, Zaher Al~Aghbari, and Ibrahim Kamel.
\newblock Detection of bots in social media: a systematic review.
\newblock {\em Information Processing \& Management}, 57(4):102250, 2020.

\bibitem{edwards2014bot}
Chad Edwards, Autumn Edwards, Patric~R Spence, and Ashleigh~K Shelton.
\newblock Is that a bot running the social media feed? testing the differences in perceptions of communication quality for a human agent and a bot agent on twitter.
\newblock {\em Computers in Human Behavior}, 33:372--376, 2014.

\bibitem{alothali2018detecting}
Eiman Alothali, Nazar Zaki, Elfadil~A Mohamed, and Hany Alashwal.
\newblock Detecting social bots on twitter: a literature review.
\newblock In {\em 2018 International conference on innovations in information technology (IIT)}, pages 175--180. IEEE, 2018.

\bibitem{Pant2025beyond}
Valeria Pantè, David Axelrod, Alessandro Flammini, Filippo Menczer, Emilio Ferrara, and Luca Luceri.
\newblock Beyond interaction patterns: Assessing claims of coordinated inter-state information operations on twitter/x.
\newblock In {\em Companion Proceedings of the ACM on Web Conference 2025}, WWW ’25, page 1234–1238. ACM, May 2025.

\bibitem{DiMarco2025}
Niccolò Di~Marco, Sara Brunetti, Matteo Cinelli, and Walter Quattrociocchi.
\newblock Post-hoc evaluation of nodes influence in information cascades: The case of coordinated accounts.
\newblock {\em ACM Transactions on the Web}, 19(2):1–19, May 2025.

\bibitem{Minici2025iohunter}
Marco Minici, Luca Luceri, Francesco Fabbri, and Emilio Ferrara.
\newblock Iohunter: Graph foundation model to uncover online information operations.
\newblock {\em Proceedings of the AAAI Conference on Artificial Intelligence}, 39(27):28258–28266, April 2025.

\bibitem{Nwala2023language}
Alexander~C. Nwala, Alessandro Flammini, and Filippo Menczer.
\newblock A language framework for modeling social media account behavior.
\newblock {\em EPJ Data Science}, 12(1), August 2023.

\bibitem{Cinus2025exposing}
Federico Cinus, Marco Minici, Luca Luceri, and Emilio Ferrara.
\newblock Exposing cross-platform coordinated inauthentic activity in the run-up to the 2024 u.s. election.
\newblock In {\em Proceedings of the ACM on Web Conference 2025}, WWW ’25, page 541–559. ACM, April 2025.

\bibitem{Cinelli2021dynamics}
Matteo Cinelli, Andraž Pelicon, Igor Mozetič, Walter Quattrociocchi, Petra~Kralj Novak, and Fabiana Zollo.
\newblock Dynamics of online hate and misinformation.
\newblock {\em Scientific Reports}, 11(1), November 2021.

\bibitem{Loru2024}
Edoardo Loru, Matteo Cinelli, Maurizio Tesconi, and Walter Quattrociocchi.
\newblock The influence of coordinated behavior on toxicity.
\newblock {\em Online Social Networks and Media}, 43–44:100289, November 2024.

\bibitem{Cresci2019capability}
Stefano Cresci, Marinella Petrocchi, Angelo Spognardi, and Stefano Tognazzi.
\newblock On the capability of evolved spambots to evade detection via genetic engineering.
\newblock {\em Online Social Networks and Media}, 9:1–16, January 2019.

\bibitem{Castellano2009}
Claudio Castellano, Santo Fortunato, and Vittorio Loreto.
\newblock Statistical physics of social dynamics.
\newblock {\em Reviews of Modern Physics}, 81(2):591–646, May 2009.

\bibitem{rainer2002opinion}
Hegselmann Rainer and Ulrich Krause.
\newblock Opinion dynamics and bounded confidence: models, analysis and simulation.
\newblock {\em Journal of Artificial Societies and Social Simulation}, 2002.

\bibitem{flache2017models}
Andreas Flache, Michael M{\"a}s, Thomas Feliciani, Edmund Chattoe-Brown, Guillaume Deffuant, Sylvie Huet, and Jan Lorenz.
\newblock Models of social influence: Towards the next frontiers.
\newblock {\em Jasss-The journal of artificial societies and social simulation}, 20(4):2, 2017.

\bibitem{das2014modeling}
Abhimanyu Das, Sreenivas Gollapudi, and Kamesh Munagala.
\newblock Modeling opinion dynamics in social networks.
\newblock In {\em Proceedings of the 7th ACM international conference on Web search and data mining}, pages 403--412, 2014.

\bibitem{adjerid2018big}
Idris Adjerid and Ken Kelley.
\newblock Big data in psychology: A framework for research advancement.
\newblock {\em American Psychologist}, 73(7):899, 2018.

\bibitem{monti2020learning}
Corrado Monti, Gianmarco De~Francisci~Morales, and Francesco Bonchi.
\newblock Learning opinion dynamics from social traces.
\newblock In {\em Proceedings of the 26th ACM SIGKDD International Conference on Knowledge Discovery \& Data Mining}, pages 764--773, 2020.

\bibitem{peralta2022opinion}
Antonio~F Peralta, J{\'a}nos Kert{\'e}sz, and Gerardo I{\~n}iguez.
\newblock Opinion dynamics in social networks: From models to data.
\newblock {\em arXiv preprint arXiv:2201.01322}, 2022.

\bibitem{Weidlich1971}
W.~Weidlich.
\newblock The statistical description of polarization phenomena in society†.
\newblock {\em British Journal of Mathematical and Statistical Psychology}, 24(2):251–266, November 1971.

\bibitem{Galam1982}
Serge Galam, Yuval Gefen~(Feigenblat), and Yonathan Shapir.
\newblock Sociophysics: A new approach of sociological collective behaviour. i. mean‐behaviour description of a strike.
\newblock {\em The Journal of Mathematical Sociology}, 9(1):1–13, November 1982.

\bibitem{Galam1991}
Serge Galam and Serge Moscovici.
\newblock Towards a theory of collective phenomena: Consensus and attitude changes in groups.
\newblock {\em European Journal of Social Psychology}, 21(1):49–74, January 1991.

\bibitem{noorazar2020classical}
Hossein Noorazar, Kevin~R Vixie, Arghavan Talebanpour, and Yunfeng Hu.
\newblock From classical to modern opinion dynamics.
\newblock {\em International Journal of Modern Physics C}, 31(07):2050101, 2020.

\bibitem{carpentras2023we}
Dino Carpentras.
\newblock Why we are failing at connecting opinion dynamics to the empirical world.
\newblock {\em Review of Artificial Societies and Social Simulations}, 2023.

\bibitem{kertesz2019algorithmic}
Janos Kertesz, Alina Sirbu, Fosca Gianotti, and Dino Pedreschi.
\newblock Algorithmic bias amplifies opinion polarization: A bounded confidence model.
\newblock In {\em StatPhys 27 Main Conference}, 2019.

\bibitem{perra2019modelling}
Nicola Perra and Luis~EC Rocha.
\newblock Modelling opinion dynamics in the age of algorithmic personalisation.
\newblock {\em Scientific reports}, 9(1):7261, 2019.

\bibitem{mei2022micro}
Wenjun Mei, Francesco Bullo, Ge~Chen, Julien~M Hendrickx, and Florian D{\"o}rfler.
\newblock Micro-foundation of opinion dynamics: Rich consequences of the weighted-median mechanism.
\newblock {\em Physical Review Research}, 4(2):023213, 2022.

\bibitem{mei2024convergence}
Wenjun Mei, Julien~M Hendrickx, Ge~Chen, Francesco Bullo, and Florian D{\"o}rfler.
\newblock Convergence, consensus and dissensus in the weighted-median opinion dynamics.
\newblock {\em IEEE Transactions on Automatic Control}, 2024.

\bibitem{bizyaeva2022nonlinear}
Anastasia Bizyaeva, Alessio Franci, and Naomi~Ehrich Leonard.
\newblock Nonlinear opinion dynamics with tunable sensitivity.
\newblock {\em IEEE Transactions on Automatic Control}, 68(3):1415--1430, 2022.

\bibitem{CLIFFORD1973}
PETER CLIFFORD and AIDAN SUDBURY.
\newblock A model for spatial conflict.
\newblock {\em Biometrika}, 60(3):581–588, 1973.

\bibitem{Holley1975}
Richard~A. Holley and Thomas~M. Liggett.
\newblock Ergodic theorems for weakly interacting infinite systems and the voter model.
\newblock {\em The Annals of Probability}, 3(4), August 1975.

\bibitem{Redner2001}
Sidney Redner.
\newblock {\em A Guide to First-Passage Processes}.
\newblock Cambridge University Press, August 2001.

\bibitem{Liggett1985}
Thomas~M. Liggett.
\newblock {\em Interacting Particle Systems}.
\newblock Springer New York, 1985.

\bibitem{DallAsta2007}
Luca Dall’Asta and Claudio Castellano.
\newblock Effective surface-tension in the noise-reduced voter model.
\newblock {\em Europhysics Letters (EPL)}, 77(6):60005, March 2007.

\bibitem{Granovsky1995}
Boris~L. Granovsky and Neal Madras.
\newblock The noisy voter model.
\newblock {\em Stochastic Processes and their Applications}, 55(1):23–43, January 1995.

\bibitem{Zillio2005}
Tommaso Zillio, Igor Volkov, Jayanth~R. Banavar, Stephen~P. Hubbell, and Amos Maritan.
\newblock Spatial scaling in model plant communities.
\newblock {\em Physical Review Letters}, 95(9), August 2005.

\bibitem{Antal2006}
T.~Antal, S.~Redner, and V.~Sood.
\newblock Evolutionary dynamics on degree-heterogeneous graphs.
\newblock {\em Physical Review Letters}, 96(18), May 2006.

\bibitem{Oborny2023}
Beáta Oborny.
\newblock Lost in translation? – caveat to the application of the voter model in ecology and evolutionary biology.
\newblock {\em Science Progress}, 106(2), April 2023.

\bibitem{peralta2021opinion}
Antonio~F Peralta, J{\'a}nos Kert{\'e}sz, and Gerardo I{\~n}iguez.
\newblock Opinion formation on social networks with algorithmic bias: dynamics and bias imbalance.
\newblock {\em Journal of Physics: Complexity}, 2(4):045009, 2021.

\bibitem{krapivsky2003dynamics}
Paul~L Krapivsky and Sidney Redner.
\newblock Dynamics of majority rule in two-state interacting spin systems.
\newblock {\em Physical Review Letters}, 90(23):238701, 2003.

\bibitem{krapivsky2021divergence}
PL~Krapivsky and S~Redner.
\newblock Divergence and consensus in majority rule.
\newblock {\em Physical Review E}, 103(6):L060301, 2021.

\bibitem{noonan2021dynamics}
James Noonan and Renaud Lambiotte.
\newblock Dynamics of majority rule on hypergraphs.
\newblock {\em Physical Review E}, 104(2):024316, 2021.

\bibitem{forgerini2024directed}
Fabricio~L Forgerini, Nuno Crokidakis, and M{\'a}rcio~AV Carvalho.
\newblock Directed propaganda in the majority-rule model.
\newblock {\em International Journal of Modern Physics C}, 35(07):2450082, 2024.

\bibitem{sah2024majority}
Ashwin Sah and Mehtaab Sawhney.
\newblock Majority dynamics: The power of one.
\newblock {\em Israel Journal of Mathematics}, pages 1--49, 2024.

\bibitem{montes2011majority}
Marco~A Montes~de Oca, Eliseo Ferrante, Alexander Scheidler, Carlo Pinciroli, Mauro Birattari, and Marco Dorigo.
\newblock Majority-rule opinion dynamics with differential latency: a mechanism for self-organized collective decision-making.
\newblock {\em Swarm Intelligence}, 5:305--327, 2011.

\bibitem{Chatterjee1977}
S.~Chatterjee and E.~Seneta.
\newblock Towards consensus: some convergence theorems on repeated averaging.
\newblock {\em Journal of Applied Probability}, 14(1):89–97, March 1977.

\bibitem{Stone1961}
M.~Stone.
\newblock The opinion pool.
\newblock {\em The Annals of Mathematical Statistics}, 32(4):1339–1342, December 1961.

\bibitem{Deffuant2000}
Guillaume Deffuant, David Neau, Frederic Amblard, and Gérard Weisbuch.
\newblock Mixing beliefs among interacting agents.
\newblock {\em Advances in Complex Systems}, 03(01n04):87–98, January 2000.

\bibitem{FORTUNATO2004}
SANTO FORTUNATO.
\newblock Universality of the threshold for complete consensus for the opinion dynamics of deffuantet al.
\newblock {\em International Journal of Modern Physics C}, 15(09):1301–1307, November 2004.

\bibitem{GMEZSERRANO2012}
JAVIER GÓMEZ-SERRANO, CARL GRAHAM, and JEAN-YVES LE~BOUDEC.
\newblock The bounded confidence model of opinion dynamics.
\newblock {\em Mathematical Models and Methods in Applied Sciences}, 22(02), February 2012.

\bibitem{Piccoli2021}
Benedetto Piccoli and Francesco Rossi.
\newblock Generalized solutions to bounded-confidence models.
\newblock {\em Mathematical Models and Methods in Applied Sciences}, 31(06):1237–1276, April 2021.

\bibitem{li2025bounded}
Grace~J Li, Jiajie Luo, and Mason~A Porter.
\newblock Bounded-confidence models of opinion dynamics with adaptive confidence bounds.
\newblock {\em SIAM Journal on Applied Dynamical Systems}, 24(2):994--1041, 2025.

\bibitem{li2024some}
Grace~Jingying Li.
\newblock {\em Some Generalizations of Bounded-Confidence Models of Opinion Dynamics}.
\newblock University of California, Los Angeles, 2024.

\bibitem{varshney2014bounded}
Kush~R Varshney.
\newblock Bounded confidence opinion dynamics in a social network of bayesian decision makers.
\newblock {\em IEEE Journal of Selected Topics in Signal Processing}, 8(4):576--585, 2014.

\bibitem{giraldez2022analyzing}
Jes{\'u}s Gir{\'a}ldez-Cru, Carmen Zarco, and Oscar Cord{\'o}n.
\newblock Analyzing the extremization of opinions in a general framework of bounded confidence and repulsion.
\newblock {\em Information Sciences}, 609:1256--1270, 2022.

\bibitem{javarone2014social}
Marco~Alberto Javarone.
\newblock Social influences in opinion dynamics: the role of conformity.
\newblock {\em Physica A: Statistical Mechanics and its Applications}, 414:19--30, 2014.

\bibitem{Valensise2023}
Carlo~M. Valensise, Matteo Cinelli, and Walter Quattrociocchi.
\newblock The drivers of online polarization: Fitting models to data.
\newblock {\em Information Sciences}, 642:119152, September 2023.

\bibitem{Baumann2020}
Fabian Baumann, Philipp Lorenz-Spreen, Igor~M. Sokolov, and Michele Starnini.
\newblock Modeling echo chambers and polarization dynamics in social networks.
\newblock {\em Physical Review Letters}, 124(4), January 2020.

\bibitem{FerrazdeArruda2022}
Henrique Ferraz~de Arruda, Felipe Maciel~Cardoso, Guilherme Ferraz~de Arruda, Alexis R.~Hernández, Luciano da~Fontoura~Costa, and Yamir Moreno.
\newblock Modelling how social network algorithms can influence opinion polarization.
\newblock {\em Information Sciences}, 588:265–278, April 2022.

\bibitem{singh2024differences}
Ashwini~Kumar Singh, Vahid Ghafouri, Jose Such, and Guillermo Suarez-Tangil.
\newblock Differences in the toxic language of cross-platform communities.
\newblock In {\em Proceedings of the International AAAI Conference on Web and Social Media}, volume~18, pages 1463--1476, 2024.

\bibitem{baqir2024news}
Anees Baqir, Alessandro Galeazzi, and Fabiana Zollo.
\newblock News and misinformation consumption: A temporal comparison across european countries.
\newblock {\em Plos one}, 19(5):e0302473, 2024.

\bibitem{saberski2024impact}
Erik Saberski, Tom Lorimer, Delia Carpenter, Ethan Deyle, Ewa Merz, Joseph Park, Gerald~M Pao, and George Sugihara.
\newblock The impact of data resolution on dynamic causal inference in multiscale ecological networks.
\newblock {\em Communications Biology}, 7(1):1442, 2024.

\bibitem{murdock2024information}
Isabel~E Murdock.
\newblock {\em Information Diffusion Over Diverse Social Media Platforms and the Simulated Cross-Platform Impacts of Interventions}.
\newblock PhD thesis, Carnegie Mellon University, 2024.

\end{thebibliography}

\end{document}